\documentclass[aps,prd,amsmath,amssymb,nofootinbib,notitlepage, 12pt]{revtex4-2}

\usepackage{stmaryrd}
\usepackage{comment}
\usepackage{color}
\usepackage{amsmath}
\usepackage{amssymb}
\usepackage{amsthm}
\usepackage{mathrsfs}
\usepackage{graphicx}
\usepackage{marvosym}
\usepackage{amsmath,bm}
\usepackage{array}
\usepackage{latexsym}
\usepackage{enumerate}
\usepackage{slashed}
\usepackage{hyperref}
\usepackage{color}
\usepackage{setspace}
\usepackage[normalem]{ulem}

\begin{document}

\title{Transverse and longitudinal spin alignment from color fields in heavy ion collisions}
\author{Di-Lun Yang}
\email{dilunyang@gmail.com}
\affiliation{Institute of Physics, Academia Sinica, Taipei, 11529, Taiwan}%
\affiliation{Physics Division, National Center for Theoretical Sciences, Taipei, 106319, Taiwan}
\begin{abstract}
We analyze the spin alignment of vector mesons stemming from spin correlation of the quark and antiquark induced by background color fields in relativistic heavy ion collisions. The quark-coalescence equation relating the collision kernel of the vector-meson kinetic equation to spin alignment is expanded to the relativistic case. Focusing on the color-octet contribution, the spin alignment for $\phi$ mesons from glasma fields with momentum dependence is investigated, where the scenarios for different spin quantization axes are considered. Moreover, we qualitatively analyze the spin alignment from isotropic color fields in the quark gluon plasma phase for comparison. In particular, we propose that the experimental measurement of spin alignment along the beam direction, dubbed as the longitudinal spin alignment, could be useful to identify the dominance of longitudinal spin correlation potentially led by the glasma effect in high-energy nuclear collisions.        	
\end{abstract}
\maketitle
\section{Introduction}
A large angular momentum can be generated in non-central heavy ion collisions, which may result in spin polarization of quark gluon plasma (QGP) through the spin-orbit interaction as theoretically proposed by Refs.~\cite{Liang:2004ph,Liang:2004xn,Voloshin:2004ha}. The polarization of QGP is expected to be inherited by the polarization of produced hadrons, $\Lambda$ hyperons in practice, which has been indeed measured by the STAR Collaboration~\cite{STAR:2017ckg,STAR:2019erd} at BNL, ALICE Collaboration at CERN~\cite{ALICE:2019onw}, and HADES Collaboration~\cite{Kornas:2020qzi}. Theoretically, the global polarization of $\Lambda$ hyperons is successfully described by the polarization spectrum induced by thermal vorticity in a global thermodynamic equilibrium  condition \cite{Becattini2013a,Becattini:2016gvu,Karpenko:2016jyx,Becattini:2017gcx,Fang:2016vpj}. However, the follow-up observations of local spin polarization with momentum dependence along the longitudinal direction by STAR \cite{STAR:2019erd} disagree with the theoretical predictions based on the global equilibrium condition \cite{Becattini:2017gcx,Xia:2018tes,Fu:2020oxj}. It is now understood that various contributions to spin polarization beyond the global equilibrium condition need to be considered. 

In particular, the so-called thermal shear corrections as one of interaction-independent contributions in local equilibrium may play a crucial role \cite{Hidaka:2017auj,Liu:2021uhn,Becattini:2021suc} for local polarization according to the hydrodynamic simulations \cite{Becattini:2021iol,Fu:2021pok,Yi:2021ryh,Florkowski:2021xvy}. Nonetheless, the numerical results from simulations are sensitive to the adopted approximations and chosen parameters \cite{Yi:2021ryh,Florkowski:2021xvy,Alzhrani:2022dpi}. In addition, further non-equilibrium contributions \cite{Hidaka:2017auj,Hidaka:2018ekt,Yang:2018lew,Shi:2020htn,Lin:2022tma,Fang:2022ttm,Fang:2023bbw,Lin:2024zik,Fang:2024vds,Banerjee:2024xnd,Lin:2024svh} or the influence from late-time magnetic fields \cite{Sun:2024isb} have been studied in recent years and a new observable, the helicity polarization along the momentum direction \cite{Becattini:2020xbh,Gao:2021rom,Yi:2021unq,Yi:2023tgg}, was proposed to distinguish different polarization mechanisms. See also Ref.~\cite{Liu:2019krs} for a far-from-equilibrium kinetic-theory approach to address the local spin polarization. Despite the complexities, most of mechanisms for spin polarization are triggered by gradient corrections on hydrodynamic variables such as the temperature, fluid velocity, and chemical potentials. It is unclear how spin polarization phenomena in heavy ion collisions could have a deeper connection to the microscopic QCD properties in extreme conditions.  

On the other hand, the spin alignment of vector mesons, characterized by the deviations of the ($00$)-component of the spin density matrix $\rho_{00}$ from its equilibrium value $1/3$~\cite{Schilling:1969um,Park:2022ayr}, have been measured in heavy ion collisions ~\cite{ALICE:2019aid,ALICE:2022dyy,Mohanty:2021vbt,STAR:2022fan}. As opposed to the relatively weak signal for spin polarization, the measured spin alignment is unexpectedly large and contradictory to the assumption of thermal equilibrium~\cite{Becattini:2007sr,Becattini:2007nd} based on the spin coalescence model \cite{Liang:2004xn,Yang:2017sdk}. It has been an open question and motivated a considerable number of theoretical studies ~\cite{Sheng:2019kmk,Sheng:2020ghv,Xia:2020tyd,Muller:2021hpe,Yang:2021fea,Goncalves:2021ziy,Sheng:2022wsy,Li:2022neh,Sheng:2022ffb,
	Li:2022vmb,Wagner:2022gza,Muller:2021hpe,Yang:2021fea,Kumar:2022ylt,Kumar:2023ghs,Sheng:2023urn,DeMoura:2023jzz,Yin:2024dnu,Dong:2023cng,Sheng:2023urn,Kumar:2023ojl}. Among the possible mechanisms, Refs.~\cite{Sheng:2019kmk,Muller:2021hpe,Yang:2021fea,Sheng:2022wsy,Sheng:2023urn,Kumar:2022ylt,Kumar:2023ghs} propose that the fluctuating strong-force fields in connection to QCD interaction could result in non-vanishing spin correlations between the quark and antiquark forming vector mesons via the quark coalescence \cite{Fries:2003vb,Greco:2003xt}. Due to fluctuating properties, such mechanisms giving rise to nonzero spin correlation will yield vanishing spin polarization for the $\Lambda$ hyperon mostly attributed to the spin polarization of the single constituted strange quark, which self-consistently explain the relatively small spin-polarization signal.  

Notably, by considering the early-time spin correlation induced by chromo-electromagnetic fields dominantly along the beam direction in the glasma phase~\cite{Lappi:2006fp,Lappi:2006hq} from the color glass condensate (CGC) effective theory~\cite{McLerran:1993ni,McLerran:1993ka,McLerran:1994vd,Gelis:2010nm,Albacete:2014fwa}, the magnitude of spin alignment for non-relativistic $\phi$ mesons is estimated in Ref.~\cite{Kumar:2023ghs}. Despite the underlying simplifications and approximations, the estimated value may roughly explain the $\rho_{00}<1/3$ of $\phi$ mesons at small transverse momenta in LHC. The presence of glasma effect is however only expected in high-energy nuclear collisions such as at the top RHIC collision energy and LHC energies. In fact, the glasma phase as a pre-equilibrium state in heavy ion collisions is usually adopted as an initial condition for phenomenological studies \cite{Schenke:2012wb}, while the existence of such a phase itself has not been experimentally confirmed. Therefore, the spin-alignment signal may also serve as an alternative probe for such a pre-equilibrium state with saturated gluons.  
On the contrary, the late-time spin correlation is proposed to be led by fluctuating vector-meson fields for $\phi$ mesons \cite{Sheng:2022wsy}, from which the global spin alignment data for $\phi$ mesons may be described in RHIC energies, whereas the strengths of such vector-meson fields cannot be estimated from a first principle. They are instead treated as fitting parameters. The alternative proposal for turbulent color fields also emphasizes the late-time spin correlation in the QGP phase~\cite{Muller:2021hpe}, whereas the order of magnitude cannot be pin downed either.  

Generically, there could exist multiple effects for spin alignment with sophisticated collision-energy, flavor, and centrality dependence. More detailed studies for spin alignment spectra depending on momenta may provide further information to disentangle different mechanisms. In this paper, we expand our previous study of the spin alignment for non-relativistic $\phi$ mesons from glasma fields or generally from background color fields in the quark coalescence scenarios \cite{Kumar:2023ghs} to the relativistic case and investigate the momentum dependence of $\rho_{00}$ with the choices of different spin quantization axes. In addition to the in-plane and out-of-plane spin alignment usually discussed in literature \cite{Sheng:2022wsy,Sheng:2023urn}, we further study the spin alignment along the beam direction. We will dub the former (in-plane and out-of-plane) and latter (beam) as the transverse and longitudinal spin alignment, respectively.  

The paper is structured as follows: In Sec.~\ref{sec:spin_transport_theory}, we briefly review the derivation of axial Wigner functions dictating the spin polarization of quarks and antiquarks from the quantum kinetic theory (QKT) with background color fields. In Sec.~\ref{sec:quark_colascence_general}, we revisit the quark coalescence scenario delineated by the collision kernel of the kinetic theory of vector mesons and derive a more general expression, especially suitable for studying the glasma effect, beyond the non-relativistic approximation and equal-mass condition. The choices of different spin quantization axes are also discussed. In Sec.~\ref{sec:spin_alignment_glasma}, we evaluate the spin alignment of $\phi$ mesons from the glasma effect and show the numerical results with momentum dependence. In Sec.~\ref{sec:spin_alignment_QGP}, we further analyze the spin alignment of $\phi$ mesons from color fields in QGP with an emphasis on qualitative features and the comparison with the results from glasma fields. Finally, we present conclusions and outlook in Sec~\ref{sec:final}. Some technical details have been relegated to the Appendix.

Throughout this paper we use 
the mostly minus signature of the Minkowski metric $\eta^{\mu\nu} = {\rm diag} (1, -1,-1,-1)  $ 
and the completely antisymmetric tensor $ \epsilon^{\mu\nu\rho\lambda} $ with $ \epsilon^{0123} = 1 $. 
We introduce the notations $A^{(\mu}B^{\nu)}\equiv A^{\mu}B^{\nu}+A^{\nu}B^{\mu}$,  $A^{[\mu}B^{\nu]}\equiv A^{\mu}B^{\nu}-A^{\nu}B^{\mu}$, and $\tilde{F}^{\mu\nu}\equiv\epsilon^{\mu\nu\alpha\beta}F_{\alpha\beta}/2$. Greek and roman indices are used for space-time and spatial components, respectively, unless otherwise specified. 

\section{Spin transport of quarks with color fields}\label{sec:spin_transport_theory}
In this section, we briefly review the QKT for spin transport of quarks with color fields constructed in Refs.~\cite{Muller:2021hpe,Yang:2021fea} and the approximate solution for dynamical spin polarization obtained in Refs.~\cite{Kumar:2023ghs,Kumar:2022ylt}. For spin transport, one may focus on the vector and axial-vector components of Wigner functions for massive fermions, $\mathcal{V}^{\mu}(p,x)$ and $\mathcal{A}^{\mu}(p,x)$, where the former and the latter are responsible for the particle-number and spin-polarization spectra, respectively. The dynamical evolution of $\mathcal{V}^{\mu}(p,x)$ and $\mathcal{A}^{\mu}(p,x)$ in phase space can then be tracked by utilizing the QKT for relativistic fermions \cite{Son:2012wh,Stephanov:2012ki,Chen:2012ca,Hidaka:2016yjf,Gao:2019znl,Weickgenannt:2019dks,Hattori:2019ahi,Wang:2019moi,Yang:2020hri,Wang:2020pej,Weickgenannt:2020aaf} (see also a recent review ~\cite{Hidaka:2022dmn} and references therein) with further inclusion of the color degrees of freedom \cite{Muller:2021hpe,Yang:2021fea,Luo:2021uog}. For quarks carrying color charges, they can be further decomposed into the color-singlet and color-octet components, 
\begin{eqnarray}
\mathcal{V}^{\mu}(p, x)=\mathcal{V}^{{\rm s}\mu}(p, x)I+\mathcal{V}^{a\mu}(p, x)\,t^a,
\quad\mathcal{A}^{\mu}(p, x)=\mathcal{A}^{{\rm s}\mu}(p, x)I+\mathcal{A}^{a\mu}(p, x)\,t^a,
\end{eqnarray}
where $t^a$ are the SU$(N_c)$ generators such that $[t^a,t^b]=if^{abc}t^c$ and $\{t^a,t^b\}=N_c^{-1}\delta^{ab}I+d^{abc}t^c$ and $I$ is the identity matrix in color space. Throughout this paper, we will use the superscripts ${\rm s}$ and $a$ to denotes the color-singlet and color-octet components for color objects, respectively. Due to the quantum nature of spin, we may adopt the power counting, $\mathcal{V}^{\mu}\sim\mathcal{O}(\hbar^0)$ and $\mathcal{A}^{\mu}\sim\mathcal{O}(\hbar)$ in the $\hbar$ expansion as the gradient expansion in phase space \cite{Yang:2020hri}, and focus on the leading-order contribution. Accordingly, for the vector Wigner functions, we have 
\begin{eqnarray}
	\mathcal{V}^{{\rm s}\mu}(p,x)=2\pi\delta(p^2-m^2)f^{\rm s}_{V}(p,x),\quad \mathcal{V}^{a\mu}(p,x)=2\pi\delta(p^2-m^2)f^{a}_{V}(p,x) 
\end{eqnarray}  
where $f^{\rm s}_{V}(p,x)$ and $f^{a}_{V}(p,x)$ denote the color-singlet and color-octet distribution functions, respectively, and $m$ represents the quark mass. For the axial-vector Wigner functions, we have
\begin{eqnarray}\nonumber
\mathcal{A}^{{\rm s}\mu}(p, x)&=&2\pi\Big[\delta(p^2-m^2)\tilde{a}^{{\rm s}\mu}(p,x)
+\frac{\hbar g}{2N_c}\delta'(p^2-m^2)p_{\nu}\tilde{F}^{a\mu\nu}f^{a}_{V}(p,x)\Big],
\\
\mathcal{A}^{a\mu}(p, x)&=&2\pi\Big[\delta(p^2-m^2)\tilde{a}^{a\mu}(p,x)
+\hbar g\delta'(p^2-m^2)p_{\nu}\Big(\tilde{F}^{a\mu\nu}f^{\rm s}_{V}(p,x)+\frac{d^{bca}}{2}\tilde{F}^{b\mu\nu}f^{c}_{V}(p,x)\Big)\Big],
\end{eqnarray}
where $g$ is the coupling constant of strong interaction, $\delta'(x)\equiv \partial\delta(x)/\partial x$, and $\tilde{F}^{a\mu\nu}\equiv \epsilon^{\mu\nu\alpha\beta}F^{a}_{\alpha\beta}/2$ with $F^{a}_{\alpha\beta}$ being the field strength of color fields. Also, $\tilde{a}^{{\rm s}\mu}(p,x)$ and $\tilde{a}^{a\mu}(p,x)$ denote the color-singlet and color-octet effective spin four vector, respectively. We may further introduce the on-shell axial Wigner functions in connection to spin polarization spectra,
\begin{eqnarray}\nonumber
	\mathcal{A}^{{\rm s}\mu}(\bm p, x)&\equiv&\int\frac{dp_0}{2\pi}\mathcal{A}^{{\rm s}\mu}(p, x)=\frac{1}{2\epsilon_{\bm p}}\Big[\tilde{a}^{{\rm s}\mu}-\frac{\hbar g}{4N_c}\tilde{F}^{a\mu\nu}\Big(\partial_{p\nu}f^{a}_{\rm V}-\frac{\epsilon_{\bm p}}{2}\partial_{p_\perp\nu}(f^{a}_{V}/\epsilon_{\bm p})\Big)\Big]_{p_0=\epsilon_{\bm p}},
	\\\nonumber
	\mathcal{A}^{a\mu}(\bm p, x)&\equiv&\int\frac{dp_0}{2\pi}\mathcal{A}^{a\mu}(p, x)
	\\\nonumber
	&=&\frac{1}{2\epsilon_{\bm p}}\Big[\tilde{a}^{ a\mu}-\frac{\hbar g}{2}\tilde{F}^{a\mu\nu}\Big(\partial_{p\nu}f^{\rm s}_{ V}-\frac{\epsilon_{\bm p}}{2}\partial_{p_\perp\nu}(f^{\rm s}_{V}/\epsilon_{\bm p})\Big)
	\\
	&&-\frac{\hbar g d^{bca}}{4}\tilde{F}^{b\mu\nu}\Big(\partial_{p\nu}f^{c}_{ V}-\frac{\epsilon_{\bm p}}{2}\partial_{p_\perp\nu}(f^{c}_{V}/\epsilon_{\bm p})\Big)\Big]_{p_0=\epsilon_{\bm p}},
\end{eqnarray}
where $\epsilon_{\bm p}\equiv\sqrt{|\bm p|^2+m^2}$. 

As constructed from the Wigner-function approach in Refs.~\cite{Muller:2021hpe,Yang:2021fea}, the dynamics of vector-charge distribution functions and effective spin four vector in phase space are governed by the coupled scalar kinetic equations (SKE), 
\begin{eqnarray}
	p^{\rho}\Big(\partial_{\rho}f^{\rm s}_V+\frac{g}{2N_c}F^a_{\nu\rho}\partial_{p}^{\nu}f^{a}_V\Big)
	=\mathcal{C}_{\rm s},
\end{eqnarray}
\begin{eqnarray}
	p^{\rho}\Big(D_{\rho}f^{a}_V+gF^a_{\nu\rho}\partial_{p}^{\nu}f^{\rm s}_V+\frac{d^{bca}}{2}gF^{b}_{\nu\rho}\partial_{p}^{\nu}f^{c}_V\Big)
	=\mathcal{C}^{a}_{\rm o},
\end{eqnarray}
and axial kinetic equations (AKE) \footnote{The $F^{a\nu\mu}\tilde{a}^{a}_{\nu}$ and $F^{a\nu\mu}\tilde{a}^{\rm s}_{\nu}$ terms were missing in Ref.~\cite{Yang:2021fea}, while they do not affect the major results therein and the related works based on the perturbative approach at weak coupling, where these terms are suppressed and neglected.},
\begin{eqnarray}
	p^{\rho}\partial_{\rho}\tilde{a}^{\rm s \mu}+\frac{g}{2N_c}\Big(p^{\rho}F^a_{\nu\rho}\partial_{p}^{\nu}\tilde{a}^{a \mu}+F^{a\nu\mu}\tilde{a}^{a}_{\nu}\Big)
	-\frac{\hbar}{4N_c}\epsilon^{\mu\nu\rho\sigma}p_{\rho}
		(D_{\sigma}gF^a_{\beta\nu})\partial_{p}^{\beta}f_{V}^a=\mathcal{C}^{\mu}_{\rm s},
\end{eqnarray}
\begin{eqnarray}\label{eq:octet_AKE}\nonumber
	&&p^{\rho}D_{\rho}\tilde{a}^{a \mu}+g\big(p^{\rho}F^a_{\nu\rho}\partial_{p}^{\nu}\tilde{a}^{\rm s \mu}+F^{a\nu\mu}\tilde{a}^{\rm s}_{\nu}\big)+\frac{d^{bca}}{2}g\big(p^{\rho}F^{b}_{\nu\rho}\partial_{p}^{\nu}\tilde{a}^{c\mu}+F^{b\nu\mu}\tilde{a}^{c}_{\nu}\big)
	\\
	&&-\frac{\hbar}{2}\epsilon^{\mu\nu\rho\sigma}p_{\rho}
		(D_{\sigma}gF^a_{\beta\nu})\partial_{p}^{\beta}f_{V}^{\rm s}=\mathcal{C}^{a\mu}_{\rm o},
\end{eqnarray}
where $D_{\rho}O^{a}\equiv\partial_{\rho}O^{a}-gf^{bca}A^b_{\rho}O^{c}$ denotes the covariant derivative for an arbitrary color object $O^a$.  
Here we do not specify the collision terms, $\mathcal{C}_{\rm s}$, $\mathcal{C}^{a}_{\rm o}$, $\mathcal{C}^{\mu}_{\rm s}$, and $\mathcal{C}^{a\mu}_{\rm o}$, explicitly, which depend on the details of interaction for the considered systems. For phenomenological purpose, we will later adopt the relaxation-time approximation for these collision terms in the QGP phase as the previous study in Ref.~\cite{Kumar:2023ghs}. See Refs.~\cite{Muller:2021hpe,Yang:2021fea} for more detailed explanations of the theoretical framework and Ref.~\cite{Kumar:2023ghs} for phenomenological applications. 

For our purpose, we may focus on $\mathcal{A}^{a\mu}(\bm p, x)$ responsible for the color-octet contribution to spin correlation. We first consider the dynamical spin polarization from $\tilde{a}^{a\mu}$ induced by space-time varying color fields obtained from Eq.~(\ref{eq:octet_AKE}). Following the perturbative approach in Ref.~\cite{Kumar:2023ghs}, one may drop the diffusion terms linear to $\tilde{a}^{{\rm s}\mu}$ and induced by color fields, whereby we approximate Eq.~(\ref{eq:octet_AKE}) as
\begin{eqnarray}
	p^{\rho}\partial_{\rho}\tilde{a}^{a \mu}-\frac{\hbar}{2}\epsilon^{\mu\nu\rho\sigma}p_{\rho}
	(\partial_{\sigma}gF^a_{\beta\nu})\partial_{p}^{\beta}f_{V}^{\rm s}=0,
\end{eqnarray}
where the collision term is also suppressed at weak coupling. One hence obtains the solution for dynamical polarization induced by color fields as  
\begin{eqnarray}\label{eq:amu_sol_no_col}
	\tilde{a}^{a\mu}(p,x)&=&\frac{\hbar g}{2p_0}\int^{x_0}_{t_{\rm i}} dx'_0\epsilon^{\mu\nu\rho\sigma}p_{\rho}\big(\partial_{x'\sigma}F^a_{\beta\nu}(x')\big)
	\partial_{p}^{\beta}f^{\rm s}_{V}(p,x')|_{\text{c}},
\end{eqnarray}
where $|_{\text{c}}=\{x^{\prime i}_{\rm T}=x^{i}_{\rm T},x^{\prime i}_{\parallel}=x^{i}_{\parallel}-p^{i}(x_0-x'_0)/p_0\}$ with $x_0$ being the present time and $t_{\rm i}\leq x_0$ denotes the initial time for having a non-vanishing integrand. Here $V^{i}_{\rm T}$ and $V^{i}_{\parallel}$ represent the perpendicular and parallel components with respect to the spatial momentum $p^{i}$ for an arbitrary spatial vector $V^{i}$, respectively.  We may express the chromo-electromagnetic fields in terms of a temporal vector $n^{\mu}=(1,\bm 0)$,
\begin{eqnarray}
	F^a_{\alpha\beta}=-\epsilon_{\mu\nu\alpha\beta}B^{a\mu}n^{\nu}+n_{\beta}E^a_{\alpha}-n_{\alpha}E^a_{\beta}.
\end{eqnarray}
By utilizing $\epsilon^{ijk}\partial_jE^a_k=\partial_0B^{ai}$ from the Bianchi identity $\partial_{\mu}\tilde{F}^{a\mu\nu}=0$ for Abelianized color fields and assuming $f^{\rm s}_{V}(p,x')=f^{\rm s}_{V}(p_0,x'_0)$ as an isotropic distribution in momentum and position space, Eq.~(\ref{eq:amu_sol_no_col}) can be written as
\begin{eqnarray}
	\tilde{a}^{ai}(p,x)&=&\frac{\hbar g}{2p_0}\int^{x_0}_{t_{\rm i}} dx'_0\big(p_0\partial_{x'0}B^{ai}(x')+\epsilon^{ijk}p_k\partial_{x'0}E^a_j(x')\big)
	\partial_{p_0}f^{\rm s}_{V}(p_0,x'_0)|_{\text{c}}.
\end{eqnarray}
When the chromo-electromagnetic fields vary rapidly only in early times for $0\leq t_{\rm i}\leq x'_0\leq t_{\rm f}$, one may neglect the temporal derivative upon $f^{\rm s}_{V}(p_0,x'_0)$ and derive
\begin{eqnarray}\nonumber
	\tilde{a}^{a i}(p,x)|_{x_0=t_{\rm f}}&\approx&-\frac{\hbar g}{2}\Big[\Big(B^{ai}(t_{\rm  i})-\frac{\epsilon^{ijk}p_{j}E^a_{k}(t_{\rm i})}{p_0}\Big)\partial_{p0}f^{\rm s}_{V}(p_0,t_{\rm i})
	\\
	&&-\Big(B^{ai}(t_{\rm f})-\frac{\epsilon^{ijk}p_{j}E^a_{k}(t_{\rm f})}{p_0}\Big)\partial_{p_0}f^{\rm s}_{V}(p_0,t_{\rm f})\Big].
\end{eqnarray}
For the glasma effect on spin alignment proposed in Ref.~\cite{Kumar:2023ghs}, one may further drop the contributions from late-time fields by assuming $B^{ai}(t_{\rm f}),\,E^{ai}(t_{\rm f})\ll B^{ai}(t_{\rm i}),\,E^{ai}(t_{\rm i})$ for time-decreasing glasma fields. We shall regard $\tilde{a}^{a\mu}(p,x)|_{t_{\rm f}=t_{\rm th}}$ as the early-time polarization induced in the glamsa phase ending at $t_{\rm th}$ as the thermalization time and the onset of the QGP phase. For $t>t_{\rm th}$, the glasma fields vanish and the collisional effect in QGP should result in the spin relaxation\footnote{The sharp transition between the glasma phase and QGP phase is assumed for simplicity, which may not reflect the realistic situation with an intermediate phase for thermalization.}. Then one has to solve 
\begin{eqnarray}
	p^{\rho}\partial_{\rho}\tilde{a}^{a\mu}=-\frac{p_0\tilde{a}^{a\mu}}{\tau^{\rm o}_{\rm R}},
\end{eqnarray}
under the relaxation-time approximation for the collision term, where $\tau^{\rm o}_{\rm R}$ represents a constant spin relaxation time characterizing the scattering effect. Here the corresponding solution reads
\begin{eqnarray}
\tilde{a}^{a\mu}(x_0)=\tilde{a}^{a\mu}(t_{\rm th})e^{-x_0/\tau^{\rm o}_{\rm R}}.
\end{eqnarray}
Neglecting the non-dynamical contribution corresponding to the part besides $\tilde{a}^{ a\mu}$ in $\mathcal{A}^{ai}$ due to the absence of color fields in late times, we arrive at 
\begin{eqnarray}\label{eq:Aa_glasma}
	\mathcal{A}^{ai}(\bm p, x)\approx -\frac{\hbar g}{4\epsilon_{\bm p}}e^{-x_0/\tau^{\rm o}_{\rm R}}\bigg[B^{ai}(t_{\rm  i})-\frac{\epsilon^{ijk}p_{j}E^a_{k}(t_{\rm i})}{\epsilon_{\bm p}}\bigg]\partial_{\epsilon_{\bm p}}f^{\rm s}_{V}(\epsilon_{\bm p},t_{\rm i})
\end{eqnarray}
for the glasma effect. Such an effect only depends on the color fields and quark distribution functions in the initial time albeit with the suppression from collisions.

On the other hand, we may instead consider the late-time effect for spin alignment stemming from chromo-electromagnetic fields in QGP as the soft thermal gluons (or turbulent color fields discussed in Ref.~\cite{Muller:2021hpe})\footnote{In fact, Ref.~\cite{Muller:2021hpe} originally focuses on the correlation of $\mathcal{A}^{{\rm s}i}(\bm p, x)$ instead of $\mathcal{A}^{ai}(\bm p, x)$. However, it was later realized that the correlator of $\mathcal{A}^{{\rm s}i}(\bm p, x)$ should be a subleading contribution for spin alignment at weak coupling, while it may lead to a leading-order contribution to the spin correlation of $\Lambda$ hyperons. }, for which we may perturbatively approximate 
\begin{eqnarray}\label{eq:Aa_QGP}
\mathcal{A}^{ai}(\bm p, x)&\approx&-\frac{\hbar g}{4\epsilon_{\bm p}}\bigg[B^{ai}(x_0)\partial_{\epsilon_{\bm p}}+\frac{\epsilon^{ijk}p_{j}E^a_{k}(x_0)}{2\epsilon_{\bm p}}\big(\partial_{\epsilon_{\bm p}}+\epsilon_{\bm p}^{-1}\big)
\bigg]f^{\rm s}_{V}(\epsilon_{\bm p},x_0),
\end{eqnarray}
where the dynamical contribution from $\tilde{a}^{a i}$ is instead dropped. Both scenarios will be further investigated in this work.

In the following sections, we will evaluate the spin alignment from the spin correlation of quarks and antiquarks induced by the correlated background color fields in the quark coalescence scenario. Due to technical complications, here we briefly summarize the general strategy and setup.
Since the spin alignment of vector mesons is measured in the rest frame of vector mesons, we solve the polarization-dependent kinetic theory for vector mesons in the rest frame with the collision term of quark coalescence shown in Ref.~\cite{Kumar:2023ghs}. The polarization dependence of such a collision term originates from the correlations of background color fields encoded in the axial Wigner functions of quarks and antiquarks obtained above. However, the correlations of color fields in the rest frame implicitly carry momentum dependence of the vector mesons in the lab frame. Such a momentum dependence is essential to yield anisotropic spin correlation for spin alignment when background color fields are isotropic in the lab frame. Moreover, the correlation functions of color fields are usually computed in the lab frame. We hence further make Lorentz transformation of the color-field correlators in the vector-meson rest frame to the lab frame in order to retrieve the momentum dependence of the spin-alignment spectrum for vector mesons. For a practical application to the spin alignment of $\phi$ mesons, we consider both the dynamical spin alignment led by anisotropic color fields from the glasma phase as an initial-state effect and the non-dynamical one triggered by isotropic color fields characterizing the interaction with soft gluons in the QGP phase as a final-state effect. In the former case, the key assumptions are first the strange quarks and antiquarks are produced in early times with strong glasma fields and second the spin relaxation time is relatively slow as opposed to the rapid thermalization of quarks and antiquarks that reach thermal equilibrium at the outset of QGP phase. For the latter case, we only consider the impact of constant background color fields on the spin correlation just before the quark coalescence for simplicity and conduct a qualitative analysis for spin alignment. 

\section{Polarized vector mesons from quark coalescence}\label{sec:quark_colascence_general}
\subsection{Generalization of the coalescence scenario}
In Ref.~\cite{Kumar:2023ghs}, the collision kernel for the quark-coalescence scenario, where a quark and an antiquark combine to form a vector meson thorough the postulated vector coupling, for the kinetic equation of vector mesons is obtained, whereas further simplification such as the non-relativistic approximation and equal mass of the quarks and antiquarks are applied to further simplify the calculation. Although the equal-mass condition is satisfied for $\phi$ mesons, the non-relativistic approximation actually violates the momentum conservation even in the rest frame of vector mesons. Such a violation is however not too severe by tuning the constituent quark mass to the value close to one half of the vector-meson mass. For accuracy and generality, in this subsection, we further calculate this collision kernel by lifting the equal-mass condition and non-relativistic approximation with particular forms of the axial Wigner functions for quarks and antiquarks.      

As shown in Ref.~\cite{Kumar:2023ghs}, the polarization dependent distribution function of vector mesons, $f_{\lambda}(q,x)$, with momentum $q$ and negligible spatial dependence obtained from the quark coalescence scenario for dilute vector mesons can be approximated as 
\begin{eqnarray}\label{eq:flambda}
	f_{\lambda}(q)\approx \frac{\Delta t}{E_q}g_{\phi}\,\mathcal{C}_{{\rm col},\lambda}(q),
\end{eqnarray}
where $g_{\phi}$ denotes the vector-coupling constant, $E_q=\sqrt{|\bm q|^2+M^2}$ with $M$ being the mass of vector mesons, and $\Delta t$ represents the coalescence time assumed to by sufficiently small. Here the subscript $\lambda=0,\pm 1$ corresponds to the polarization of the spin-one vector mesons. The normalized $00$-th component of the spin density matrix of vector mesons that characterizes its spin alignment can be obtained from   
\begin{eqnarray}\label{eq:rho00_def}
	\rho_{00}(q)=\frac{\int d\Sigma\cdot qf_{0}(q,x)}{\int d\Sigma\cdot q\big(f_{0}(q,x)+f_{+1}(q,x)+f_{-1}(q,x)\big)},
\end{eqnarray}
where $\Sigma^{\mu}$ corresponds to a freeze-out hypersurface.   
The collision kernel for quark coalescence reads
\begin{eqnarray}
\mathcal{C}_{{\rm col},\lambda}(q)=\int \frac{d^3k}{(2\pi)^3}\delta\big(q_0-\epsilon_{q}(\bm q/2+\bm k)-\epsilon_{\bar{q}}(\bm q/2-\bm k)\big)\tilde{\mathcal{C}}_{{\rm col},\lambda}(q,k),
\end{eqnarray}
where $\epsilon_{q/\bar{q}}(\bm p)\equiv\sqrt{|\bm p|^2+m^2_{q/\bar{q}}}$ with $m_q$ and $m_{\bar{q}}$ being the mass of quarks and of antiquarks, respectively, and
\begin{eqnarray}\nonumber\label{eq:SigmaVA}
	&&\tilde{\mathcal{C}}_{{\rm col},\lambda}(q,k)
	\\\nonumber
	&&=N_{m}
	{\rm Tr_{c}}\Bigg\{\Bigg[\frac{ f_{Vq}(\bm p,x)f_{V\bar{q}}(\bm p',x)}{\epsilon_{q}(\bm p)\epsilon_{\bar{q}}(\bm p')}
	-\frac{4}{m_{q}m_{\bar{q}}}\mathcal{A}_{q}(\bm p,x)\cdot \mathcal{A}_{\bar{q}}(\bm p',x)\Bigg]\Bigg(1-\frac{2|k\cdot\epsilon(\lambda, {\bm q})|^2}{N_m}\Bigg)
	\\\nonumber
	&&\quad -\frac{4}{m_{q}m_{\bar{q}}}\Bigg[2{\rm Re}\Big(\epsilon(\lambda,{\bm q})\cdot\mathcal{A}_{q}(\bm p,x)\epsilon^*(\lambda,{\bm q})\cdot\mathcal{A}_{\bar{q}}(\bm p',x)\Big)
	+\frac{2}{N_m}
	\Big(2k\cdot\mathcal{A}_{q}(\bm p,x)k\cdot\mathcal{A}_{\bar{q}}(\bm p',x)
	\\\nonumber
	&&\quad
	+2{\rm Re}\big(k\cdot\epsilon(\lambda,{\bm q})\mathcal{A}_{q}(\bm p,x)\cdot \epsilon^*(\lambda,{\bm q})\big)k\cdot\mathcal{A}_{\bar{q}}(\bm p',x)
	\\
	&&\quad+2{\rm Re}\big(k\cdot\epsilon(\lambda,{\bm q})\mathcal{A}_{\bar{q}}(\bm p',x)\cdot \epsilon^*(\lambda,{\bm q})\big)k\cdot\mathcal{A}_{q}(\bm p,x)
	\Big)\Bigg]
	\Bigg\}\Bigg|_{p,p'}
\end{eqnarray}
is explicitly expressed in terms the axial Wigner functions of quarks and antiquarks. 
Here $|_{p,p'}=\{p=\frac{q}{2}+k,p'=\frac{q}{2}-k\}$ for $p$ and $p'$ being the momenta of quarks and antiquarks, respectively. Also, $\epsilon(\lambda,{\bm q})$ denotes the polarization four-vector for vector mesons and
\begin{eqnarray}\label{eq:def_Nm}
	N_{m}=\frac{1}{2}\big(M^2-(m_{q}+m_{\bar{q}})^2\big).
\end{eqnarray}
We also use subscripts $q$ and $\bar{q}$ to represent the contributions from quarks and antiquarks for the vector distribution functions and axial-vector Wigner functions.
We then work in the vector-meson rest frame, $q^{\mu}=(M,{\bm 0})$, such that $\epsilon^{\mu}(\lambda,0)=(0,\bm\epsilon_{\lambda})\equiv\epsilon_{\lambda}^{\mu}$. Accordingly, the onshell and kinematic conditions give rise to
\begin{eqnarray}
	k_0=\frac{m_q^2-m_{\bar{q}}^2}{2M},\quad |\bm k|= \tilde{k}\equiv \frac{1}{2M}\big[(m_q^2-m_{\bar{q}}^2)^2+M^4-2M^2(m_q^2+m_{\bar{q}}^2)\big]^{1/2},
\end{eqnarray} 
and
\begin{eqnarray}
\delta\big(q_0-\epsilon_{q}(\bm q/2+\bm k)-\epsilon_{\bar{q}}(\bm q/2-\bm k)\big)
=\frac{\big(M^4-(m_q^2-m_{\bar{q}}^2)^2\big)
\delta\big(|\bm k|-\tilde{k}\big)}{8M^3\tilde{k}}.
\end{eqnarray}
In Ref.~\cite{Kumar:2023ghs}, the equal-mass condition $m_q=m_{\bar{q}}$ and non-relativistic approximation $M-(m_q+m_{\bar{q}})\ll M$ are further applied to simplify the explicit form of Eq.~(\ref{eq:SigmaVA}) for the $k$ integral. 

In order to conduct the $k$ integral without using the non-relativistic approximation, we instead introduce explicit forms of $\mathcal{A}_{q/\bar{q}}^{i}(\bm p,x)$ and $\mathcal{A}_{q/\bar{q}}^{i}(\bm p',x)$ based on the possible structures induced by color fields in light of Eqs.~(\ref{eq:Aa_glasma}) and (\ref{eq:Aa_QGP}), the polarization along chromo-magnetic fields and the one perpendicular to the momentum and chromo-electric fields as the spin Hall effect. Let us now assume
\begin{eqnarray}\nonumber
&&\mathcal{A}_{q/\bar{q}}^{i}(\bm p,x)|_{\bm q=0}=\mathcal{A}_{q/\bar{q}}^{(0)i}(|\bm k|,x)+k_{j}\mathcal{A}_{q/\bar{q}}^{(1)ij}(|\bm k|,x),\quad
f_{Vq/\bar{q}}(\bm p,x)|_{\bm q=0}=f_{Vq/\bar{q}}(|\bm k|,x),
\\
&&\mathcal{A}_{q/\bar{q}}^{i}(\bm p',x)|_{\bm q=0}=\mathcal{A}_{q/\bar{q}}^{(0)i}(|\bm k|,x)-k_{j}\mathcal{A}_{q/\bar{q}}^{(1)ij}(|\bm k|,x),\quad
f_{Vq/\bar{q}}(\bm p',x)|_{\bm q=0}=f_{Vq/\bar{q}}(|\bm k|,x),
\end{eqnarray}
with $\mathcal{A}_{q/\bar{q}}^{(1)ji}(|\bm k|,x)=-\mathcal{A}_{q/\bar{q}}^{(1)ij}(|\bm k|,x)$. On the other hand, due to the constraint $p_{\mu}\mathcal{A}_{q/\bar{q}}^{\mu}(\bm p,x)=0$, we also have the relations,
\begin{eqnarray}\nonumber
\mathcal{A}_{q/\bar{q}}^{0}(\bm p,x)|_{\bm q=0}&=&-\Big(p_i\mathcal{A}_{q/\bar{q}}^{i}(\bm p,x)/p_0\Big)\Big|_{\bm q=0}=-\epsilon_{q}(\bm k)^{-1}k_i\mathcal{A}_{q/\bar{q}}^{i}(\bm p,x)|_{\bm q=0},
\\
\mathcal{A}_{q/\bar{q}}^{0}(\bm p',x)|_{\bm q=0}&=&-\Big(p'_i\mathcal{A}_{q/\bar{q}}^{i}(\bm p',x)/p'_0\Big)\Big|_{\bm q=0}=\epsilon_{\bar{q}}(\bm k)^{-1}k_i\mathcal{A}_{q/\bar{q}}^{i}(\bm p',x)|_{\bm q=0}.
\end{eqnarray}
Note that $\mathcal{A}_{q/\bar{q}}^{0}(\bm p,x)|_{\bm q=0}$ and $\mathcal{A}_{q/\bar{q}}^{0}(\bm p',x)|_{\bm q=0}$ do not contain the contributions from $\mathcal{A}_{q/\bar{q}}^{(1)ij}(|\bm k|,x)$ due to its anti-symmetric property.
We may then apply the relations,
\begin{eqnarray}\label{eq:useful_rel_1}
	k^ik^j\rightarrow |\bm k|^2\Big[\bar{z}^2\epsilon^{i}_{\lambda}\epsilon^{j}_{\lambda}-\frac{(1-\bar{z}^2)}{2}\hat{\Theta}^{ij}_{\lambda}\Big],\quad \hat{\Theta}^{ij}_{\lambda}=\eta^{ij}+\epsilon^{i}_{\lambda}\epsilon^{j}_{\lambda},\quad
	\bar{z}=\frac{-k\cdot\epsilon_{\lambda}}{|\bm k|},
\end{eqnarray}
and
\begin{eqnarray}\nonumber\label{eq:useful_rel_2}
&&k^ik^jk^lk^n\rightarrow |\bm k|^4\hat{\Theta}^{ijln}_{\lambda},
\\\nonumber
&&\hat{\Theta}^{ijln}_{\lambda}=\Big[\bar{z}^2\epsilon^{i}_{\lambda}\epsilon^{j}_{\lambda}-\frac{(1-\bar{z}^2)}{2}\hat{\Theta}^{ij}_{\lambda}\Big]\Big[\bar{z}^2\epsilon^{l}_{\lambda}\epsilon^{n}_{\lambda}-\frac{(1-\bar{z}^2)}{2}\hat{\Theta}^{ln}_{\lambda}\Big]+(i\leftrightarrow l)+(i\leftrightarrow n),
\\
&&\hat{\Theta}^{ijln}_{\lambda}\epsilon_{\lambda i}\epsilon_{\lambda j}=3\bar{z}^4\epsilon^{l}_{\lambda}\epsilon^{n}_{\lambda}-\frac{\bar{z}^2(1-\bar{z}^2)}{2}\hat{\Theta}^{ln}_{\lambda}=-\frac{\bar{z}^2(1-\bar{z}^2)}{2}\eta^{ln}_{\lambda}
-\frac{\bar{z}^2(1-7\bar{z}^2)}{2}\epsilon^{l}_{\lambda}\epsilon^{n}_{\lambda},
\end{eqnarray}
for $\epsilon_{\lambda}$ being real and drop the terms with odd powers of $\bm k$ when evaluating the integral. 

Based on the complicated yet straightforward computations shown in the Appendix, it is found
\begin{eqnarray}\nonumber\label{eq:f_lambda_general}
	f_{\lambda}&=& \frac{\Delta t}{E_q}g_{\phi}N_mN_c^2C_V\bigg\{1-\frac{1}{\pi^2 m_qm_{\bar{q}}C_VN_c^2}\bigg[\bigg(1-2C_m+\frac{\tilde{k}^2}{2\epsilon_{q\tilde{k}}\epsilon_{\bar{q}\tilde{k}}}\Big(1-\frac{2C_m}{3}\Big)+\frac{4k_0^2}{3\epsilon_{q\tilde{k}}\epsilon_{\bar{q}\tilde{k}}}\bigg)
	\eta^{ij}\big\langle\mathcal{A}_{qi}^{a(0)}\mathcal{A}_{\bar{q}j}^{a(0)}\big\rangle
	\\\nonumber
	&&+\bigg(2\Big(1-\frac{4C_m}{3}\Big)-\frac{16C_m\tilde{k}^2}{15\epsilon_{q\tilde{k}}\epsilon_{\bar{q}\tilde{k}}}\bigg){\rm Re}\Big(\big\langle\epsilon_{\lambda}\cdot\mathcal{A}_{q}^{a(0)}\epsilon_{\lambda}\cdot \mathcal{A}_{\bar{q}}^{a(0)}\big\rangle\Big)
	+\frac{|\bm k|^2}{3}\Big(1-\frac{2C_m}{5}\Big)\big\langle\mathcal{A}_{qij}^{a(1)}\mathcal{A}_{\bar{q}}^{a(1)ij}\big\rangle
	\\
	&&+\frac{|\bm k|^2}{6}\Big(1+\frac{32C_m}{5}\Big){\rm Re}\Big(\big\langle\epsilon^{i}_{\lambda}\epsilon^{*j}_{\lambda}\mathcal{A}_{qil}^{a(1)}\mathcal{A}_{\bar{q}j}^{a(1)l}\big\rangle\Big)
	\bigg]\bigg\}_{|\bm k|=\tilde{k}},
\end{eqnarray}
where 
\begin{eqnarray}\label{eq:def_CV}
C_V=\frac{1}{2\pi^2}\Big(1-\frac{2C_m}{3}\Big)\frac{f_{Vq}(\epsilon_{q\tilde{k}},x_0)f_{V\bar{q}}(\epsilon_{\bar{q}\tilde{k}},x_0)}{\epsilon_{q\tilde{k}}\epsilon_{\bar{q}\tilde{k}}}\Big|_{|\bm k|=\tilde{k}}, 
\end{eqnarray}
\begin{eqnarray}
	\epsilon_{q/\bar{q}\tilde{k}}\equiv\sqrt{\tilde{k}^2+m^2_{q/\bar{q}}}=\frac{1}{2M}\big[(m_{q/\bar{q}}^2-m_{\bar{q}/q}^2)+M^2\big],
\end{eqnarray}
and  
\begin{eqnarray}\label{eq:k2_rel}
	C_{m}\equiv\frac{\tilde{k}^2}{N_m}=\frac{1}{2}\left(1-\frac{(m_q-m_{\bar{q}})^2}{M^2}\right).
\end{eqnarray}
The notation $\langle O^a\,O^b\rangle$ represents a quantum expectation value of the correlation function for field operators $O^a$ and $O^b$.
Here we have dropped $f^{a}_{Vq,\bar{q}}$-related terms as the higher-order corrections for the polarization independent part and simply denoted $f_{Vq,\bar{q}}=f^{\rm s}_{Vq,\bar{q}}$ for brevity.
Despite the assumption for a particular form of the axial Wigner functions involved, Eq.~(\ref{eq:f_lambda_general}) now serves as a general expression satisfying the momentum conservation and applicable for $m_q\neq m_{\bar{q}}$. 

\subsection{Applications to glasma fields}
We will now apply the formalism derived above to the color-octet contribution for the spin correlation from glasma fields in Eq.~(\ref{eq:Aa_glasma}). Now, taking $p^{\mu}=q^{\mu}/2+k^{\mu}$ in the vector-meson rest frame, one could read out
\begin{eqnarray}\label{eq:Aq0i_Bfield}
	\mathcal{A}_{q}^{a(0)i}(|\bm k|,x)&=&-\frac{\hbar g}{4\epsilon_{\bm k}}e^{-x_0/\tau^{\rm o}_{\rm R}}B^{ai}(0)\partial_{\epsilon_{\bm k}}f_{Vq}(\epsilon_{\bm k},0),
	\\\label{eq:Aq1ij_Efield}
	\mathcal{A}_{q}^{a(1)ik}(|\bm k|,x)&=&-\frac{\hbar g\epsilon^{ijk}}{4\epsilon_{\bm k}^2}e^{-x_0/\tau^{\rm o}_{\rm R}}E^a_j(0)\partial_{\epsilon_{\bm k}}f_{Vq}(\epsilon_{\bm k},0),
\end{eqnarray}
where we have taken $t_{\rm i}=0$ for simplicity. 
Similar equations for $\mathcal{A}_{a\bar{q}}^{(0)i}(|\bm k|,x)$ and $\mathcal{A}_{a\bar{q}}^{(1)ik}(|\bm k|,x)$ with overall minus signs due to opposite color charges can be found. However, there exists a subtlety for the decomposition of $\mathcal{A}_{q}^{ai}({\bm p},x)|_{\bm q=0}$ into $\mathcal{A}_{q}^{a(0)i}(|\bm k|,x)$ and $\mathcal{A}_{q/\bar{q}}^{a(1)ik}(|\bm k|,x)$ above. In fact, the color fields here implicitly incorporate the momentum dependence. For instance, $B^{ai}(0)$ in Eq.~(\ref{eq:Aq0i_Bfield}) should be explicitly written as $B^{ai}(0, x^i_{\rm T}, x^i_{\parallel}-k^it_{\rm th}/p_0)$, which implicitly depends on $k^i$, with $p_0=M/2+k_0$, where $x^i_{\rm T}$ and $x^i_{\parallel}$ denote the transverse and parallel components with $\bm k$, respectively. Similarly, for $\mathcal{A}_{a\bar{q}}^{(0)i}(|\bm k|,x)$, we will have the contribution from $B^{ai}(0, x^i_{\rm T}, x^i_{\parallel}+k^it_{\rm th}/p_0)$. 
Nevertheless, we have to compute the color-field correlation to obtain the spin correlation. Based on the translational invariance, the color-field correlator like $\langle B^{ai}(0,\bm x)B^{ai}(0,\bm y)\rangle$ should only depend on $|\bm x-\bm y|\propto|\bm k|$. Therefore, the application of Eqs.~(\ref{eq:Aq0i_Bfield}) and (\ref{eq:Aq1ij_Efield}) to Eq.~(\ref{eq:f_lambda_general}) is still valid\footnote{In general, when considering $\big\langle B^{ai}(x_0,\bm x+\bm k t^{\rm th}/p_+)B^{aj}(y_0,\bm x-\bm k t^{\rm th}/p_-)\big\rangle$ or $\big\langle B^{ai}(0,\bm x+\bm k t^{\rm th}/p_+)B^{aj}(0,\bm y-\bm k t^{\rm th}/p_-)\big\rangle$, the correlations no longer just depend on $|\bm k|$.}.

Denoting $f_{Vq/\bar{q}}(\epsilon_{q/\bar{q}\bm k},0)=f^{(0)}_{Vq/\bar{q}}(\epsilon_{q/\bar{q}\bm k})$ for short and inserting  Eqs.~(\ref{eq:Aq0i_Bfield}) and (\ref{eq:Aq1ij_Efield}) into Eq.~(\ref{eq:f_lambda_general}), we obtain
\begin{eqnarray}\nonumber
	f_{\lambda}&=& \frac{\Delta t}{E_q}g_{\phi}N_mN_c^2C_V\bigg\{1-\frac{\hbar^2g^2C_A}{\pi^2 m_qm_{\bar{q}}C_VN_c^2}\bigg[C_{B1}\sum_{j=x,y,z}\chi_B^{jj}
	+C_{B2}{\rm Re}\Big(\epsilon_{\lambda i}\epsilon^*_{\lambda j}\chi_B^{ij}\Big)
	\\
	&&	+C_{E1}\sum_{j=x,y,z}\chi_E^{jj}
	+C_{E2}{\rm Re}\big(\epsilon_{\lambda i}\epsilon^*_{\lambda j}\chi^{ij}_{E}\big)\bigg]\bigg\}_{|\bm k|=\tilde{k}},
\end{eqnarray}
where 
\begin{eqnarray}\label{eq:CA}
	C_A=\frac{(\partial_{\epsilon_{q\tilde{k}}}f^{(0)}_{qV})(\partial_{\epsilon_{\bar{q}\tilde{k}}}f^{(0)}_{\bar{q}V})}{16\epsilon_{q\tilde{k}}\epsilon_{\bar{q}\tilde{k}}} e^{-2x_0/\tau^{\rm o}_{\rm R}},
\end{eqnarray}
and 
\begin{eqnarray}\label{eq:def_chiBE}\nonumber
	\chi_B^{ij}&=& \big\langle B^{ai}(0,\bm x+\bm k t^{\rm th}/p_+)B^{aj}(0,\bm x-\bm k t^{\rm th}/p_-)\big\rangle,
	\\
	\chi_E^{ij}&=&\big\langle E^{ai}(0,\bm x+\bm k t^{\rm th}/p_+)E^{aj}(0,\bm x-\bm k t^{\rm th}/p_-)\big\rangle,
\end{eqnarray}
as functions of $|\bm k|=\tilde{k}$ and $p_{\pm}=(M^2\pm (m_q^2-m_{\bar{q}}^2))/(2M)$. Here we have utilized
\begin{eqnarray}
	\big\langle\mathcal{A}_{qij}^{a(1)}\mathcal{A}_{\bar{q}}^{a(1)ij}\big\rangle=-\frac{2C_A}{\epsilon_{q\tilde{k}}\epsilon_{\bar{q}\tilde{k}}}\sum_{j=x,y,z}\chi^{jj}_{E}
\end{eqnarray}
and
\begin{eqnarray}
	{\rm Re}\Big(\big\langle\epsilon^{i}_{\lambda}\epsilon^{*j}_{\lambda}\mathcal{A}_{qil}^{a(1)}\mathcal{A}_{\bar{q}j}^{a(1)l}\big\rangle\Big)=\frac{C_A}{\epsilon_{q\tilde{k}}\epsilon_{\bar{q}\tilde{k}}}\Big(\sum_{j=x,y,z}\chi^{jj}_{E}-{\rm Re}\big(\epsilon_{\lambda i}\epsilon^*_{\lambda j}\chi^{ij}_{E}\big)\Big).
\end{eqnarray}
The related coefficients explicitly read
\begin{eqnarray}\nonumber
&&C_{B1}=1-2C_m+\frac{\lambda_{\tilde{k}}}{2}\Big(1-\frac{2C_m}{3}\Big)+\frac{4\lambda_{k_0}}{3},
\quad C_{B2}=-2\Big(1-\frac{4C_m}{3}\Big)+\frac{16C_m\lambda_{\tilde{k}}}{15},
\\\nonumber
&&C_{E1}=-\frac{1}{2}\Big(1-\frac{8C_m}{3}\Big)\lambda_{\tilde{k}},
\quad C_{E2}=-\frac{1}{6}\Big(1+\frac{32C_m}{5}\Big)\lambda_{\tilde{k}},
\end{eqnarray}
where
\begin{eqnarray}\nonumber
	\lambda_{\tilde{k}}\equiv\frac{\tilde{k}^2}{\epsilon_{q\tilde{k}}\epsilon_{\bar{q}\tilde{k}}}&=&
	\frac{\big(M^2-(m_q+m_{\bar{q}})^2\big)\big(M^2-(m_q-m_{\bar{q}})^2\big)}{M^4-(m_q^2-m_{\bar{q}}^2)^2},
	\\
	\lambda_{k_0}\equiv\frac{k_0^2}{\epsilon_{q\tilde{k}}\epsilon_{\bar{q}\tilde{k}}}&=&
	\frac{(m_q^2-m_{\bar{q}}^2)^2}{M^4-(m_q^2-m_{\bar{q}}^2)^2}.
\end{eqnarray}
When taking $M\approx m_{q}+m_{\bar{q}}$ and $m_{q}\approx m_{\bar{q}}$, it is found $\lambda_{\tilde{k}}\approx 0$ and $\lambda_{k_0}\approx 0$ along with $C_{m}\approx 1/2$, respectively, which correspond to the approximation and condition adopted in Ref.~\cite{Kumar:2023ghs}.

\subsection{Spin quantization axes}
Spin alignment characterized by $\rho_{00}$ depends on the choice of a spin quantization axis. In heavy ion collision, it is usually chosen to be the direction perpendicular to the reaction plane as the direction along the total angular momentum,  known as the "out-of-plane" spin alignment. For the coordinates in this paper, we will assign $\hat{z}$ as the beam direction and $\hat{y}$ as the angular-momentum direction perpendicular to the reaction plane. 
Consequently, by setting 
\begin{eqnarray}
	\bm\epsilon_0=(0,1,0),\quad \bm\epsilon_{+1}=-\frac{1}{\sqrt{2}}(i,0,1),\quad \bm\epsilon_{-1}=\frac{1}{\sqrt{2}}(-i,0,1),
\end{eqnarray}
we derive
\begin{eqnarray}\nonumber
	f_{0}&=& \frac{\Delta t}{E_q}g_{\phi}N_mN_c^2C_V\bigg[1-\frac{\hbar^2g^2C_A}{\pi^2 m_qm_{\bar{q}}C_VN_c^2}\big(C_{B1}\sum_{j=x,y,z}\chi_B^{jj}
	+C_{B2}\chi_B^{yy}
	\\
	&&+C_{E1}\sum_{j=x,y,z}\chi_E^{jj}
	+C_{E2}\chi^{yy}_{E}\big)\bigg]_{|\bm k|=\tilde{k}}
\end{eqnarray}
and
\begin{eqnarray}\nonumber
	f_{\pm 1}&=& \frac{\Delta t}{E_q}g_{\phi}N_mN_c^2C_V\bigg[1-\frac{\hbar^2g^2C_A}{\pi^2 m_qm_{\bar{q}}C_VN_c^2}\bigg(C_{B1}\sum_{j=x,y,z}\chi_B^{jj}
	+\frac{C_{B2}}{2}(\chi_B^{xx}+\chi_B^{zz})
	\\
	&&	+C_{E1}\sum_{j=x,y,z}\chi_E^{jj}
	+\frac{C_{E2}}{2}(\chi^{xx}_{E}+\chi^{zz}_{E})\bigg)\bigg]_{|\bm k|=\tilde{k}}.
\end{eqnarray}
In principle, when calculating $\rho_{00}(q)$ from Eq.~(\ref{eq:rho00_def}), we have to integrate $f_{\lambda}$ over $d\Sigma\cdot q$, while the integral leads to just a trivial overall prefactor when neglecting the spatial dependence of $f_{\lambda}$. Nevertheless, $f_{\lambda}$ still depend on time and we may take $x_0=t_{\rm eq}$ as the time at the freeze-out hypersurface for chemical (or more precisely spin) equilibrium.  
Consequently, the out-of-plane spin alignment is given by
\begin{eqnarray}\label{eq:rho00_QKT_global}\nonumber
 	\rho^{\rm o}_{00}(q)
 	&=&\frac{1-C_g\big[\big(C_{B1}+C_{B2}\big)\chi^{yy}_{B}+C_{B1}(\chi^{xx}_{B}+\chi^{zz}_{B}\big)+\big(C_{E1}+C_{E2}\big)\chi^{yy}_{E}+C_{E1}(\chi^{xx}_{E}+\chi^{zz}_{E}\big)\big]_{|\bm k|=\tilde{k}}}{3-C_g\big[\big(3C_{B1}+C_{B2}\big)\chi_{B}+\big(3C_{E1}+C_{E2}\big)\chi_{E}\big]_{|\bm k|=\tilde{k}}}
 	\\
\end{eqnarray}
from Eq.~(\ref{eq:rho00_def}), where
\begin{eqnarray}
C_g=\frac{\hbar^2g^2C_A}{\pi^2 m_qm_{\bar{q}}C_VN_c^2}\Big|_{x_0=t_{\rm eq}},
\quad
\chi_{B,E}=\sum_{j=x,y,z}\chi^{jj}_{B,E}.
\end{eqnarray}
For small spin correlations, $\chi^{jj}_{B}(t^{\rm th})\ll 1$, $\rho^{\rm o}_{00}(q)$ reduces to
\begin{eqnarray}
\rho^{\rm o}_{00}(q)\approx \frac{1}{3}\bigg[1+\frac{1}{3}C_g\Big(C_{B2}\big(\chi^{xx}_{B}+\chi^{zz}_{B}-2\chi^{yy}_{B}\big)
+C_{E2}\big(\chi^{xx}_{E}+\chi^{zz}_{E}-2\chi^{yy}_{E}\big)\Big)
\bigg]_{|\bm k|=\tilde{k}}.
\end{eqnarray}

Furthermore, when choosing the helicity frame such that the spin quantization axis is along the momentum direction of the vector meson, we can also read out the expression of $\rho^{\rm h}_{00}$ from Eq.~(\ref{eq:rho00_QKT_global}),
\begin{eqnarray}\nonumber
	\rho^{\rm h}_{00}(q)
	&=&\frac{1-C_g\big[\big(C_{B1}+C_{B2}\big)\chi^{\hat{q}}_{B}+C_{B1}(\chi_{B}-\chi^{\hat{q}}_{B}\big)+\big(C_{E1}+C_{E2}\big)\chi^{\hat{q}}_{E}+C_{E1}(\chi_{E}-\chi^{\hat{q}}_{E}\big)\big]_{|\bm k|=\tilde{k}}}{3-C_g\big[\big(3C_{B1}+C_{B2}\big)\chi_{B}+\big(3C_{E1}+C_{E2}\big)\chi_{E}\big]_{|\bm k|=\tilde{k}}},
	\\
\end{eqnarray}
where $\chi^{\hat{q}}_{B,E}=\hat{q}_{i}\hat{q}_{j}\chi^{ij}_{B,E}$.
The expression in the helicity frame is obtained by making the replacements, $\chi^{yy}_{B,E}\rightarrow \chi^{\hat{q}}_{B,E}$ and $\chi^{xx}_{B,E}+\chi^{zz}_{B,E}\rightarrow \chi_{B,E}-\chi^{\hat{q}}_{B,E}$, in Eq.~(\ref{eq:rho00_QKT_global}). Notably, when we average the momentum directions by taking $\chi^{\hat{q}}_{B,E}\rightarrow \chi_{B,E}/3$, it is found $\rho^{\rm h}_{00}(q)\rightarrow 1/3$.

On the other hand, one may alternatively choose the beam direction as the spin quantization axis. In such a case, one can simply make the replacements, $\chi^{yy}_{B,E}\rightarrow\chi^{zz}_{B,E}$, $\chi^{zz}_{B,E}\rightarrow\chi^{xx}_{B,E}$, and $\chi^{xx}_{B,E}\rightarrow\chi^{yy}_{B,E}$ in Eq.~(\ref{eq:rho00_QKT_global}). We hence obtain
\begin{eqnarray}\nonumber\label{eq:rho00_QKT_beam}
	\rho^{\rm b}_{00}(q)
	&=&\frac{1-C_g\big[\big(C_{B1}+C_{B2}\big)\chi^{zz}_{B}+C_{B1}(\chi^{xx}_{B}+\chi^{yy}_{B}\big)+\big(C_{E1}+C_{E2}\big)\chi^{zz}_{E}+C_{E1}(\chi^{xx}_{E}+\chi^{yy}_{E}\big)\big]_{|\bm k|=\tilde{k}}}{3-C_g\big[\big(3C_{B1}+C_{B2}\big)\chi_{B}+\big(3C_{E1}+C_{E2}\big)\chi_{E}\big]_{|\bm k|=\tilde{k}}}
	\\
\end{eqnarray}
as the master equation for the "longitudinal" spin alignment. 
Analogously, we may also consider the spin quantization axis along the $\hat{x}$ direction as for the "in-plane" spin alignment, for which we make the replacements, $\chi^{yy}_{B,E}\rightarrow\chi^{xx}_{B,E}$, $\chi^{zz}_{B,E}\rightarrow\chi^{yy}_{B,E}$, and $\chi^{xx}_{B,E}\rightarrow\chi^{zz}_{B,E}$ in Eq.~(\ref{eq:rho00_QKT_global}). It is found
\begin{eqnarray}\nonumber\label{eq:rho00_QKT_inplane}
	\rho^{\rm i}_{00}(q)
	=\frac{1-C_g\big[\big(C_{B1}+C_{B2}\big)\chi^{xx}_{B}+C_{B1}(\chi^{yy}_{B}+\chi^{zz}_{B}\big)+\big(C_{E1}+C_{E2}\big)\chi^{xx}_{E}+C_{E1}(\chi^{yy}_{E}+\chi^{zz}_{E}\big)\big]_{|\bm k|=\tilde{k}}}{3-C_g\big[\big(3C_{B1}+C_{B2}\big)\chi_{B}+\big(3C_{E1}+C_{E2}\big)\chi_{E}\big]_{|\bm k|=\tilde{k}}}.
	\\
\end{eqnarray}

\section{Spin alignment from glasma}\label{sec:spin_alignment_glasma}
\subsection{Correlations of glasma fields}
We may now estimate the spin alignment induced by color fields from the glasma following the similar setup in Ref.~\cite{Kumar:2023ghs} to evaluate $\chi^{ij}_{B/E}$. 
From now on, we set $\hbar=1$.
In such a case, only the correlations between longitudinal color fields parallel to the beam direction exist, which take the form \cite{Guerrero-Rodriguez:2021ask}
\begin{eqnarray}\label{eq:BEfields_early}
	\langle E^{az}(t,x_{\perp})E^{az}(t,y_{\perp})\rangle_{t=0}&\approx&\frac{1}{2}g^2N_c(N_c^2-1)
	\Omega_{+}(x_{\perp},y_{\perp}),
	\\
	\label{eq:Bz_corr}
	\langle B^{az}(t,x_{\perp})B^{az}(t,y_{\perp})\rangle_{t=0}&\approx&\frac{1}{2}g^2N_c(N_c^2-1)
	\Omega_{-}(x_{\perp},y_{\perp}),
\end{eqnarray}
where we use the subscripts $\perp$ here to represent the transverse components. 
We may further adopt the Golec-Biernat W\"usthoff (GBW) dipole distribution such that \cite{Golec-Biernat:1998zce,Guerrero-Rodriguez:2021ask}
\begin{align}\label{eq:GBW}
	\Omega_{\pm}(u_{\perp},v_{\perp})=\Omega(u_{\perp},v_{\perp})=\frac{Q_s^4}{g^4N_c^2}\left(\frac{1-e^{-Q_s^2|u_{\perp}-v_{\perp}|^2/4}}{Q_s^2|u_{\perp}-v_{\perp}|^2/4}\right)^2,
\end{align}
where $Q_s$ denotes the saturation momentum. Accordingly, it is found
\begin{eqnarray}\label{eq:chiBE_zz}
	\chi^{zz}_{B}=\chi^{zz}_{E}=\chi_{\rm GBW}\equiv \frac{Q_s^4(N_c^2-1)}{2g^2N_c}\left(\frac{1-e^{-Q_s^2\tilde{k}^2(t^{\rm th}\lambda_m)^2/4}}{Q_s^2\tilde{k}^2(t^{\rm th}\lambda_m)^2/4}\right)^2,
\end{eqnarray} 
where $\lambda_m=|p_+^{-1}+p_-^{-1}|=4M^3/(M^4-(m_q^2-m_{\bar{q}}^2)^2)$. Now, due to the dominance of longitudinal color fields and Eq.~(\ref{eq:chiBE_zz}), Eq.~(\ref{eq:rho00_QKT_global}) reduces to
\begin{eqnarray}
	\rho^{\rm o}_{00}(q)
	&\approx&\frac{1-C_g\big[C_{B1}+C_{E1}\big]\chi^{zz}_{B}}{3-C_g\big[\big(3C_{B1}+C_{B2}\big)+\big(3C_{E1}+C_{E2}\big)\big]\chi^{zz}_{B}}.
\end{eqnarray}

Except for the color-field correlators, $\rho^{\rm o}_{00}(q)$ also depends on the quark and antiquark distribution functions at the initial time and the freeze-out hypersurface. As adopted in Ref.~\cite{Kumar:2023ghs}, we postulate
\begin{eqnarray}\label{eq:noneq_fq}
f^{(0)}_{Vq/\bar{q}}(\epsilon_{q/\bar{q}\bm k})=\Big({\rm exp}\big(\sqrt{|\bm k|^2+m^2_{q/\bar{q}}}/Q_s\big)+1\Big)^{-1}
\end{eqnarray} 
as the non-equilibrium distribution functions of quarks and antiquakrs in the glasma state and hence $\partial_{\epsilon_{q/\bar{q}\bm k}}f^{(0)}_{Vq/\bar{q}}=-Q_s^{-1}f^{(0)}_{Vq/\bar{q}}(1-f^{(0)}_{Vq/\bar{q}})$ for $C_A$ in Eq.~(\ref{eq:CA}). On the other hand, the quark and antiquark distribution functions, $f_{Vq/\bar{q}}(\epsilon_{q/\bar{q}\bm k},t_{\rm th})$ in $C_V$, come from those in the QGP phase at late times. Consequently, one may take
\begin{eqnarray}
f_{Vq/\bar{q}}(\epsilon_{q/\bar{q}\bm k},t_{\rm th})=\Big({\rm exp}\big(\sqrt{|\bm k|^2+m^2_{q/\bar{q}}}/T_{\rm eq}\big)+1\Big)^{-1}
\end{eqnarray}
as the equilibrium distribution functions with the freeze-out temperature $T_{\rm eq}$ to evaluate $C_V$ in Eq.~(\ref{eq:def_CV}). In such a case, the competition between the spin-relaxation factor $e^{-2t_{\rm eq}/\tau^{\rm o}_{\rm R}}$ and the ratio of $f^{(0)}_{Vq/\bar{q}}(\epsilon_{q/\bar{q}\bm k})$ to $f_{Vq/\bar{q}}(\epsilon_{q/\bar{q}\bm k},t_{\rm th})$ also becomes crucial for the magnitude of spin alignment.

\subsection{Retrieving momentum dependence}
As also discussed in Ref.~\cite{Kumar:2023ghs}, to retrieve the momentum dependence of $\rho_{00}$, we could rewrite $B^{ai}_{\rm r}$ and $E^{ai}_{\rm r}$ as the color fields in the rest frame in terms of those in the lab frame through 
\begin{eqnarray}\nonumber
	B^{ai}_{\rm r}&=&\gamma (B^{ai}+\epsilon^{ijk}v_jE^a_{k})-(\gamma-1){\bm B}^a\cdot\hat{{\bm v}}\hat{v}^i,
	\\
	E^{ai}_{\rm r}&=&\gamma (E^{ai}-\epsilon^{ijk}v_jB^a_{k})-(\gamma-1){\bm E}^a\cdot\hat{{\bm v}}\hat{v}^i,
\end{eqnarray} 
where $\gamma=1/\sqrt{1-|{\bm v}|^2}$ with $v^i=q^i/\sqrt{|\bm q|^2+M^2}$ and $\hat{v}^i=v^i/|\bm v|$. By dropping the correlations between a chromo-magnetic field and an electric one and those between color fields along different directions, we accordingly find
\begin{eqnarray}\label{eq:BrBr_general}
	\langle B^{ai}_{\rm r}B^{ai}_{\rm r}\rangle&=&\gamma^2\big(\langle B^{ai}B^{ai}\rangle +\epsilon^{ijk}v_j\epsilon^{ij'k'}v_{j'}\langle E^{a}_kE^{a}_{k'}\rangle\big)
	\\\nonumber
	&&-2\gamma(\gamma-1)\langle B^{ai}B^{ai}\rangle\hat{v}_i^2+(\gamma-1)^2\hat{v}_i^2\sum_{j=x,y,z}\langle B^{aj}B^{aj}\rangle
\end{eqnarray}	
and
\begin{eqnarray}\label{eq:ErEr_general}\nonumber
	\langle E^{ai}_{\rm r}E^{ai}_{\rm r}\rangle&=&\gamma^2\big(\langle E^{ai}E^{ai}\rangle +\epsilon^{ijk}v_j\epsilon^{ij'k'}v_{j'}\langle B^{a}_kB^{a}_{k'}\rangle\big)
	\\
	&&-2\gamma(\gamma-1)\langle E^{ai}E^{ai}\rangle\hat{v}_i^2+(\gamma-1)^2\hat{v}_i^2\sum_{j=x,y,z}\langle E^{aj}E^{aj}\rangle.
\end{eqnarray}	
For brevity, we do not show the spatial dependence of all the equal-time correlators at $t=0$ above explicitly\footnote{In principle, under the Lorentz transformation, the space-time coordinates of the color fields in the lab frame read
	$t=\gamma(t_{\rm r}+{\bm v}\cdot{\bm x}_{\rm r})$, ${\bm x}_{\rm L}=\gamma({\bm x}_{\rm r, L}+{\bm v}t_{\rm r})$, and ${\bm x}_{\rm T}={\bm x}_{\rm r, T}$,
	where the subscripts, $\rm T$ and $\rm L$, correspond to the transverse and longitudinal components with respect to $\bm v$. However, we may assume the color fields and quarks are present at the initial time $t=0$ albeit the initial $t_{\rm r}$ depends on $\bm v$. On the other hand, the same spatial shift for two color fields does not affect the correlator by imposing the spatial translation invariance. Based on our assumptions, the color fields in the lab frame do not depend on $\bm v$.}. 
Considering the dominance of longitudinal chromo-fields in the early time, we approximate
\begin{eqnarray}\nonumber
	\langle B^{ax}_{\rm r}B^{ax}_{\rm r}\rangle&\approx&\gamma^2v_y^2\langle E^{az}E^{az}\rangle
	+(\gamma-1)^2\hat{v}_x^2\langle B^{az}B^{az}\rangle,
	\\\nonumber
	\langle B^{ay}_{\rm r}B^{ay}_{\rm r}\rangle&\approx&\gamma^2v_x^2\langle E^{az}E^{az}\rangle
	+(\gamma-1)^2\hat{v}_y^2\langle B^{az}B^{az}\rangle,
	\\
	\langle B^{az}_{\rm r}B^{az}_{\rm r}\rangle&\approx&\big(\gamma^2+(1-\gamma^2)\hat{v}_z^2\big)\langle B^{az}B^{az}\rangle, 
\end{eqnarray}	
and
\begin{eqnarray}\nonumber
	\langle E^{ax}_{\rm r}E^{ax}_{\rm r}\rangle&\approx&\gamma^2v_y^2\langle B^{az}B^{az}\rangle
	+(\gamma-1)^2\hat{v}_x^2\langle E^{az}E^{az}\rangle,
	\\\nonumber
	\langle E^{ay}_{\rm r}E^{ay}_{\rm r}\rangle&\approx&\gamma^2v_x^2\langle B^{az}B^{az}\rangle
	+(\gamma-1)^2\hat{v}_y^2\langle E^{az}E^{az}\rangle,
	\\
	\langle E^{az}_{\rm r}E^{az}_{\rm r}\rangle&\approx&\big(\gamma^2+(1-\gamma^2)\hat{v}_z^2\big)\langle E^{az}E^{az}\rangle.
\end{eqnarray}	
When adopting the GBW distribution, we have $\langle B^{az}B^{az}\rangle=\langle E^{az}E^{az}\rangle$ and accordingly
\begin{eqnarray}\nonumber
	\chi^{xx}_B(q)&=&\chi^{xx}_E(q)=\big(\gamma^2v_y^2+(\gamma-1)^2\hat{v}_x^2\big)\chi_{\rm GBW},
	\\\nonumber
	\chi^{yy}_B(q)&=&\chi^{yy}_E(q)=\big(\gamma^2v_x^2+(\gamma-1)^2\hat{v}_y^2\big)\chi_{\rm GBW},
	\\
	\chi^{zz}_B(q)&=&\chi^{zz}_E(q)=\big(\gamma^2+(1-\gamma^2)\hat{v}_z^2\big)\chi_{\rm GBW},
\end{eqnarray}	
where $\chi_{\rm GBW}$ is introduced in Eq.~(\ref{eq:chiBE_zz}). From Eq.~(\ref{eq:rho00_QKT_global}), we hence obtain
\begin{eqnarray}\nonumber
	\rho^{\rm o}_{00}(q)
	&\approx&\frac{1-\big[\big(\gamma^2v_x^2+(\gamma-1)^2\hat{v}_y^2\big)(C_{BE1}+C_{BE2})+\big(\gamma^2v_y^2+(\gamma-1)^2\hat{v}_x^2+\gamma^2+(1-\gamma^2)\hat{v}_z^2\big)C_{BE1}\big]}{3-\big[\big(\gamma^2v_y^2+(\gamma-1)^2\hat{v}_x^2+\gamma^2v_x^2+(\gamma-1)^2\hat{v}_y^2+\gamma^2+(1-\gamma^2)\hat{v}_z^2\big)(3B_{BE1}+C_{BE2})\big]}
	\\
	&=&\frac{1-\gamma^2C_{BE1}-C^{\rm o}_x\hat{v}_x^2-C^{\rm o}_y\hat{v}_y^2-C^{\rm o}_z\hat{v}_z^2}{3-\gamma^2(3C_{BE1}+C_{BE2})-C_x\hat{v}_x^2-C_y\hat{v}_y^2-C_z\hat{v}_z^2},
\end{eqnarray}
where $C_{BE1,2}=C_g(C_{B1,2}+C_{E1,2})\chi_{\rm GBW}|_{|\bm k|=\tilde{k}}$,
\begin{eqnarray}\nonumber
C^{\rm o}_x&=&\gamma^2v^2(C_{BE1}+C_{BE2})+(\gamma-1)^2C_{BE1},
\\\nonumber
C^{\rm o}_y&=&(\gamma-1)^2(C_{BE1}+C_{BE2})+\gamma^2v^2C_{BE1},
\\
C^{\rm o}_z&=&(1-\gamma^2)C_{BE1},
\end{eqnarray}
and
\begin{eqnarray}\nonumber
	C_x&=&C_y=\big((\gamma-1)^2+\gamma^2v^2\big)(3C_{BE1}+C_{BE2}),
	\\
	C_z&=&(1-\gamma^2)(3C_{BE1}+C_{BE2}).
\end{eqnarray}

One may examine the asymptotic forms of $\rho^{\rm o}_{00}$ for small and large momenta. In the small-momentum limit, $\gamma\rightarrow 1$ or $v\rightarrow 0$, we have 
\begin{eqnarray}\nonumber
	&&C^{\rm o}_x\approx v^2(C_{BE1}+C_{BE2}),\quad  C^{\rm o}_y\approx v^2C_{BE1},\quad C^{\rm o}_z\approx -v^2C_{BE1},
	\\
	&&C_x=C_y\approx v^2(3C_{BE1}+C_{BE2}),\quad 	C_z\approx -v^2(3C_{BE1}+C_{BE2}),
\end{eqnarray}
which leads to
\begin{eqnarray}
	\rho^{\rm o}_{00}(q)
	\approx \frac{1-C_{BE1}-v^2\big(C_{BE1}(1+\hat{v}_x^2+\hat{v}_y^2-\hat{v}_z^2)+ C_{BE2}\hat{v}_x^2\big)}{3-(3C_{BE1}+C_{BE2})-v^2(3C_{BE1}+C_{BE2})(1+\hat{v}_x^2+\hat{v}_y^2-\hat{v}_z^2)}.
\end{eqnarray}
With small correlations and using $\hat{v}_x^2+\hat{v}_y^2+\hat{v}_z^2=1$,  $\rho^{\rm o}_{00}(q)$ further reduces to
\begin{eqnarray}
	\rho^{\rm o}_{00}(q)
	\approx \frac{1}{3}\bigg(1+\frac{C_{BZ2}}{3}\big(1+2v_y^2-v_x^2\big)\bigg),
\end{eqnarray}
which matches the qualitative analysis in Ref.~\cite{Kumar:2023ghs}.
In the large-momentum limit, $\gamma\gg 1$, we have 
\begin{eqnarray}\nonumber
	C^{\rm o}_x&\approx&\gamma^2(2C_{BE1}+C_{BE2})-2
	\gamma C_{BE1}+\mathcal{O}(\gamma^0),
	\\\nonumber
	C^{\rm o}_y&\approx & \gamma^2(2C_{BE1}+C_{BE2})-2
	\gamma(C_{BE1}+C_{BE2})+\mathcal{O}(\gamma^0),
	\\
	C^{\rm o}_z&\approx&-\gamma^2C_{BE1}+\mathcal{O}(\gamma^0),
\end{eqnarray}
and
\begin{eqnarray}\nonumber
	C_x&=&C_y\approx 2\gamma^2(3C_{BE1}+C_{BE2})-2\gamma(3C_{BE1}+C_{BE2})+\mathcal{O}(\gamma^0),
	\\
	C_z&\approx&-\gamma^2(3C_{BE1}+C_{BE2})+\mathcal{O}(\gamma^0),
\end{eqnarray}
which give rise to
\begin{eqnarray}\nonumber
	\rho^{\rm o}_{00}(q)
	&\approx&\frac{1-\gamma^2\big(C_{BE1}+(2C_{BE1}+C_{BE2})(\hat{v}_x^2+\hat{v}_y^2)-C_{BE1}\hat{v}_z^2\big)
		+2\gamma\big((C_{BE1}+C_{BE2})\hat{v}_y^2+C_{BE1}\hat{v}_x^2\big)}{3-\gamma^2\big((3C_{BE1}+C_{BE2})(1+2\hat{v}_x^2+2\hat{v}_y^2-\hat{v}_z^2\big)+2\gamma(3C_{BE1}+C_{BE2})(\hat{v}_x^2+\hat{v}_y^2)}
	\\
	&\approx& \frac{1-\gamma^2(3C_{BE1}+C_{BE2})(1-\hat{v}_z^2\big)+2\gamma\big(C_{BE1}(1-\hat{v}_z^2)+C_{BE2}\hat{v}_y^2\big)}{3-3\gamma^2(3C_{BE1}+C_{BE2})(1-\hat{v}_z^2\big)+2\gamma(3C_{BE1}+C_{BE2})(1-\hat{v}_z^2)}
	,
\end{eqnarray}
where we have utilized $v^2=1-1/\gamma^2$ and dropped the higher-order term in the large-$\gamma$ expansion to derive the final result up to the leading-order correction on spin alignment. For the weak-correlation limit, $\rho_{00}(q)$ above further reduces to
\begin{eqnarray}\label{eq:rho00o_large_gamma}
\rho^{\rm o}_{00}(q)\approx \frac{1}{3}\bigg[1+2\gamma\bigg(C_{BE1}(1-\hat{v}_z^2)+C_{BE2}\hat{v}_y^2-\frac{(3C_{BE1}+C_{BE2})(1-\hat{v}_z^2)}{3}\bigg)\bigg].
\end{eqnarray} 

On the other hand, we may also consider the spin quantization axis along the beam direction. Based on Eq.~(\ref{eq:rho00_QKT_beam}) with the GBW distribution,  we have
\begin{eqnarray}\nonumber
	\rho^{\rm b}_{00}(q)
	&\approx&\frac{1-\big[\big(\gamma^2v_y^2+(\gamma-1)^2\hat{v}_x^2+\gamma^2v_x^2+(\gamma-1)^2\hat{v}_y^2\big)C_{BE1}+\big(\gamma^2+(1-\gamma^2)\hat{v}_z^2\big)(C_{BE1}+C_{BE2})\big]}{3-\big[\big(\gamma^2v_y^2+(\gamma-1)^2\hat{v}_x^2+\gamma^2v_x^2+(\gamma-1)^2\hat{v}_y^2+\gamma^2+(1-\gamma^2)\hat{v}_z^2\big)(3B_{BE1}+C_{BE2})\big]}
	\\
	&=&\frac{1-\gamma^2(C_{BE1}+C_{BE2})-C^{\rm b}_x\hat{v}_x^2-C^{\rm b}_y\hat{v}_y^2-C^{\rm b}_z\hat{v}_z^2}{3-\gamma^2(3C_{BE1}+C_{BE2})-C_x\hat{v}_x^2-C_y\hat{v}_y^2-C_z\hat{v}_z^2},
\end{eqnarray}
where
\begin{eqnarray}\nonumber\label{eq:Cbxyz_coeff}
	C^{\rm b}_x&=&C^{\rm b}_y=\big((\gamma-1)^2+\gamma^2v^2\big)C_{BE1},
	\\
	C^{\rm b}_z&=&(1-\gamma^2)(C_{BE1}+C_{BE2}).
\end{eqnarray}
In the small-momentum limit, $\gamma\rightarrow 1$ or $v\rightarrow 0$, we have 
\begin{eqnarray}
	C^{\rm b}_x= C^{\rm b}_y\approx v^2C_{BE1},\quad 	C^{\rm b}_z\approx -v^2(C_{BE1}+C_{BE2}),
\end{eqnarray}
and accordingly
\begin{eqnarray}
	\rho^{\rm b}_{00}(q)
	&\approx&\frac{1-(C_{BE1}+C_{BE2})-v^2\big(C_{BE1}(1+\hat{v}_x^2+\hat{v}_y^2-\hat{v}_z^2)+ C_{BE2}(1-\hat{v}_z^2)\big)}{3-(3C_{BE1}+C_{BE2})-v^2(3C_{BE1}+C_{BE2})(1+\hat{v}_x^2+\hat{v}_y^2-\hat{v}_z^2)},
\end{eqnarray}
which further reduces to
\begin{eqnarray}
	\rho^{\rm b}_{00}(q)
	\approx \frac{1}{3}\bigg(1-\frac{C_{BE2}}{3}(2+v_x^2+v_y^2)\bigg)
\end{eqnarray}
with small correlations. In the large-momentum limit, $\gamma\gg 1$, we have 
\begin{eqnarray}
	C^{\rm b}_x=C^{\rm b}_y\approx\big(\gamma^2-2\gamma\big)C_{BE1}+\mathcal{O}(\gamma^0),\quad
		C^{\rm b}_z\approx-\gamma^2(C_{BE1}+C_{BE2})+\mathcal{O}(\gamma^0),
\end{eqnarray}
and hence
\begin{eqnarray}\nonumber\label{eq:rho00b_large_gamma}
	\rho^{\rm b}_{00}(q)
	&\approx&\frac{1-\gamma^2(1-\hat{v}_z^2)(C_{BE1}+C_{BE2})-\big(\gamma^2-2\gamma\big)C_{BE1}(\hat{v}_x^2+\hat{v}_y^2)}{3-\gamma^2\big((3C_{BE1}+C_{BE2})(1+2\hat{v}_x^2+2\hat{v}_y^2-\hat{v}_z^2\big)+2\gamma(3C_{BE1}+C_{BE2})(\hat{v}_x^2+\hat{v}_y^2)}
	\\
	&\approx&\frac{1-\gamma^2(2C_{BE1}+C_{BE2})(1-\hat{v}_z^2)}{3-3\gamma^2(3C_{BE1}+C_{BE2})(1-\hat{v}_z^2\big)},
\end{eqnarray}
which further reduces to
\begin{eqnarray}
	\rho^{\rm b}_{00}(q)
	\approx \frac{1}{3}\big(1+\gamma^2C_{BE1}(1-\hat{v}_z^2)\big)
\end{eqnarray}
with small correlations.

Finally, we consider the in-plane spin alignment. From Eq.~(\ref{eq:rho00_QKT_inplane}) with the GBW distribution, it is found
\begin{eqnarray}\nonumber
	\rho^{\rm i}_{00}(q)
	&\approx&\frac{1-\big[\big(\gamma^2v_y^2+(\gamma-1)^2\hat{v}_x^2\big)(C_{BE1}+C_{BE2})+\big(\gamma^2v_x^2+(\gamma-1)^2\hat{v}_y^2+\gamma^2+(1-\gamma^2)\hat{v}_z^2\big)C_{BE1}\big]}{3-\big[\big(\gamma^2v_y^2+(\gamma-1)^2\hat{v}_x^2+\gamma^2v_x^2+(\gamma-1)^2\hat{v}_y^2+\gamma^2+(1-\gamma^2)\hat{v}_z^2\big)(3B_{BE1}+C_{BE2})\big]}
	\\
	&=&\frac{1-\gamma^2C_{BE1}-C^{\rm i}_x\hat{v}_x^2-C^{\rm i}_y\hat{v}_y^2-C^{\rm i}_z\hat{v}_z^2}{3-\gamma^2(3C_{BE1}+C_{BE2})-C_x\hat{v}_x^2-C_y\hat{v}_y^2-C_z\hat{v}_z^2},
\end{eqnarray}
where
\begin{eqnarray}\nonumber
	C^{\rm i}_x&=&C^{\rm o}_y=(\gamma-1)^2(C_{BE1}+C_{BE2})+\gamma^2v^2C_{BE1},
	\\\nonumber
	C^{\rm i}_y&=&C^{\rm o}_x=\gamma^2v^2(C_{BE1}+C_{BE2})+(\gamma-1)^2C_{BE1},
	\\
	C^{\rm i}_z&=&C^{\rm o}_z=(1-\gamma^2)C_{BE1}.
\end{eqnarray}
One may also analyze the small- and large-momentum limits, while the results are same as those for the out-of-plane spin alignment by exchanging the $x$ and $y$ indices.

\subsection{Numerical results}
For phenomenological study, the four momentum of vector mesons can be written as
\begin{eqnarray}
q_{\mu}=(m_{T}\cosh y_{q},q_{T}\cos\phi,q_{T}\sin\phi,m_T\sinh y_{q}),
\end{eqnarray}
where $m_T=\sqrt{M^2+q_{T}^2}$ with $q_T=\sqrt{q_x^2+q_y^2}$ and $y_{q}=\frac{1}{2}\ln[(E_q+q_z)/(E_q-q_z)]$ as the momentum rapidity with $E_q=\sqrt{|\bm q|^2+M^2}=m_{T}\cosh y_{q}$. The velocity of vector mesons can be expressed as functions of $(m_{T},\phi,y_q)$,
\begin{eqnarray}
v_x=\frac{q_T\cos\phi}{m_{T}\cosh y_{q}},\quad v_y=\frac{q_T\sin\phi}{m_{T}\cosh y_{q}},\quad v_z=\tanh y_{q},\quad
|\bm v|=\frac{\sqrt{m_{T}^2\cosh^2 y_{q}-M^2}}{m_{T}\cosh y_{q}},
\end{eqnarray} 
which further yields $\gamma=m_T\cosh y_q/M$. Since both the transverse and longitudinal spin alignment here only depend on $\hat{v}_i^2$ for $i=x,y,z$, the results are symmetric for $y_q\leftrightarrow -y_q$.
We will now focus on the spin alignment for $\phi$ mesons. Following Ref.~\cite{Kumar:2023ghs}, we take $m_q=m_{
\bar{q}}\approx 500$ MeV as the constituent quark mass for strange quarks, $M=1.019$ GeV for $\phi$ mesons, $T_{\rm eq}\approx 150$ MeV as the freeze-out temperature at chemical equilibrium, and $t_{\rm th}=0.2$ fm as the thermalization time at the end of the glasma phase. We also adopt $(\tau^{\rm o}_{\rm R})^{-1}\approx 0.04$ GeV~\footnote{The value of $\tau^{\rm o}_{\rm R}$ was originally obtained from the spin relaxation time defined in Ref.~\cite{Hongo:2022izs} based on the non-relativistic approximation and numerically estimated in Ref.~\cite{Kumar:2023ghs}, while it is coincidentally close to the extracted value from the phenomenological study \cite{Banerjee:2024xnd}. In general, we postulate that the spin relaxation is much slower than the charge or momentum relaxation for strange quarks.}  and $t_{\rm eq}\approx 5$ fm as the chemical-equilibrium time at the freeze-out hypersurface as the same setup in Ref.~\cite{Kumar:2023ghs}. Notably, with the setup above, one numerically finds
\begin{eqnarray}\label{eq:CBE12_numerical}
C_{BE1}>0,\quad C_{BE2}<0,\quad |C_{BE2}|\gg C_{BE1}.
\end{eqnarray}
Furthermore, for more direct comparison with experimental observations, one should in principle calculate $\rho_{00}(q)$ weighted by the momentum spectrum carrying additional information such as the collision energies and centrality. See e.g., Refs.~\cite{Sheng:2022wsy,Sheng:2023urn,Kumar:2023ojl}. For simplicity and self-consistency of the adopted approximations, we do not include the weighting. 

\begin{figure}[t]
	\includegraphics[scale=0.65]{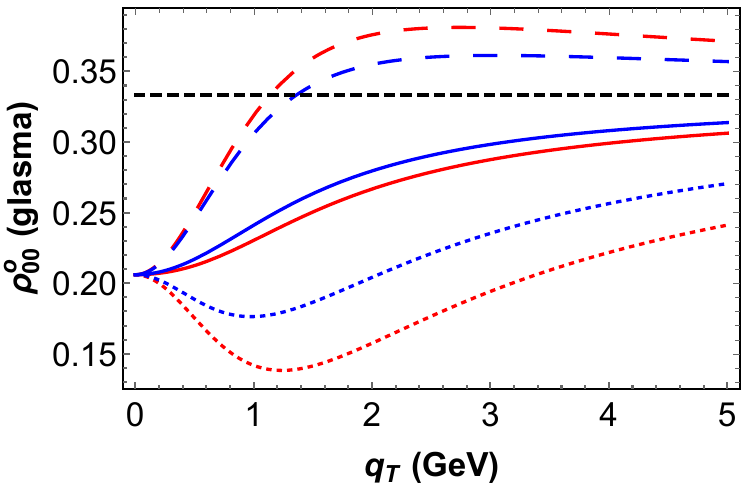}\quad\includegraphics[scale=0.65]{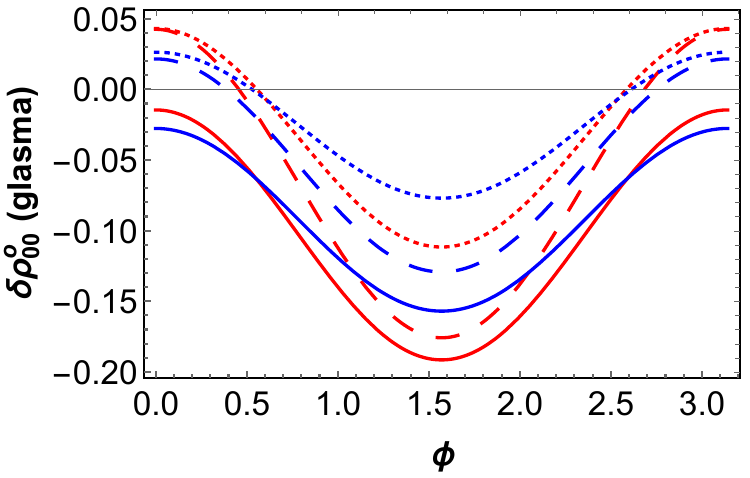}
	
	\caption{The left panel shows $\rho^{\rm o}_{00}$ with the transverse-momentum dependence, where the red and blue correspond to $y_q=0$ and $y_q=1$, respectively. The solid, dashed, and dotted curves are for $\phi=\pi/4$, $\phi=0$, and $\phi=\pi/2$. The black dashed line represents the base line for $\rho^{\rm o}_{00}=1/3$. The right panel shows $\delta\rho^{\rm o}_{00}=\rho^{\rm o}_{00}-1/3$ with the azimuthal angle dependence, where the red and blue correspond to $y_q=0$ and $y_q=1$, respectively. The solid, dashed, and dotted curves are for $q_T=1$, $q_T=2$, and $q_T=4$ GeV. \label{fig:rho00_out_qT_phi}}
\end{figure}
\begin{figure}[t]
	\includegraphics[scale=0.65]{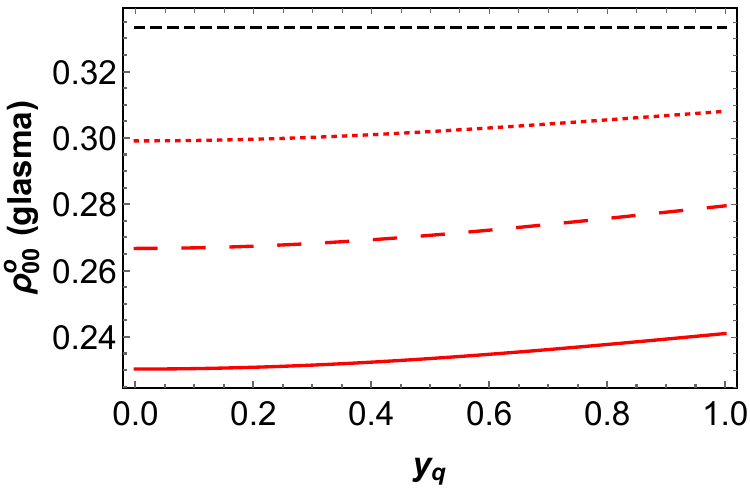}\quad\includegraphics[scale=0.65]{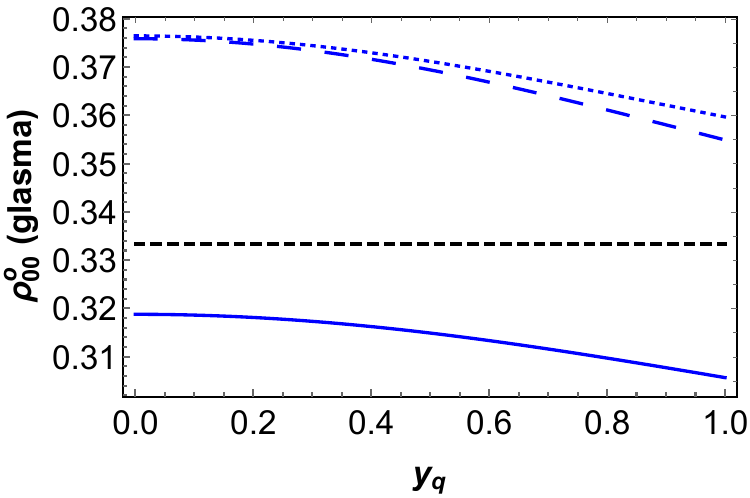}
	
	\caption{The left and right panels show $\rho^{\rm o}_{00}$ for $\phi$ mesons with the momentum-rapidity dependence for $\phi=\pi/4$ (red) and $\phi=0$ (blue), respectively. The solid, dashed, and dotted curves are for $q_T=1$, $q_T=2$, and $q_T=4$ GeV. The black dashed lines represent the base lines for $\rho^{\rm o}_{00}=1/3$.  \label{fig:rho00_out_qT_yq}}
\end{figure}
\begin{figure}[t]
	\includegraphics[scale=0.5]{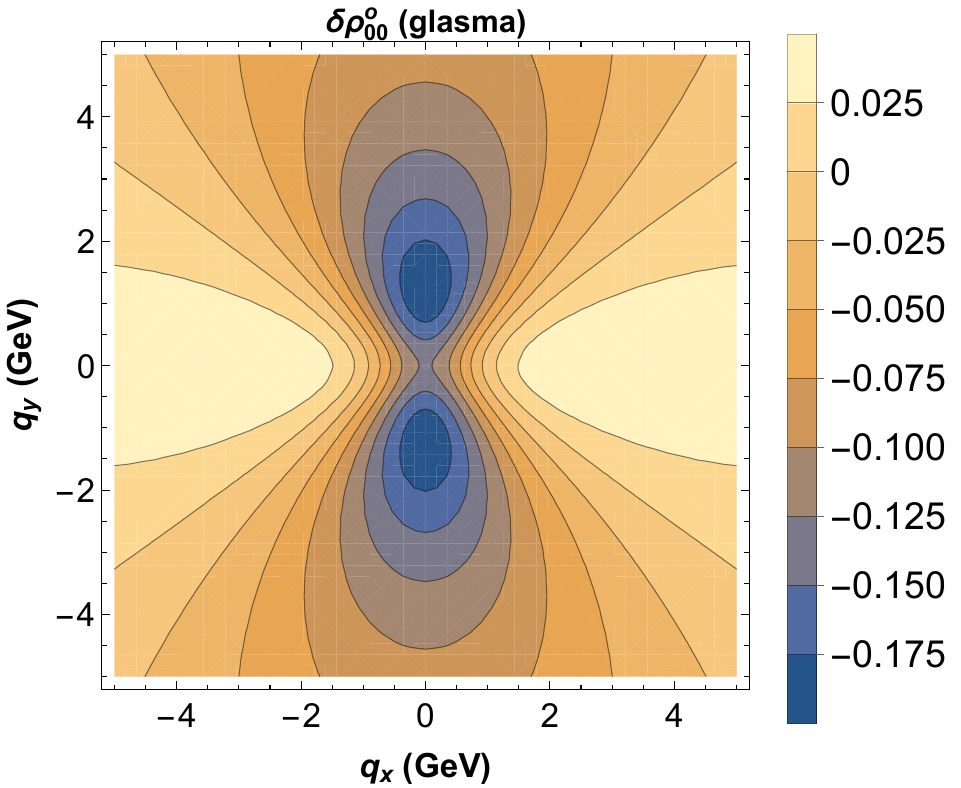}\quad\includegraphics[scale=0.5]{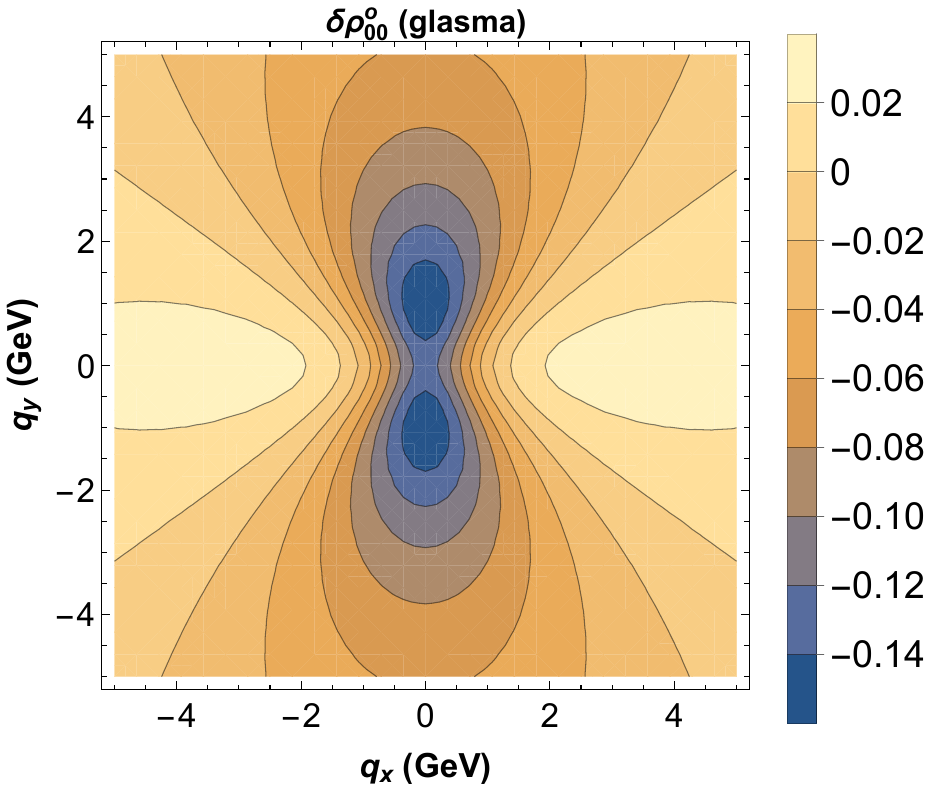}
	
	\caption{The contour plots for $\delta\rho^{\rm o}_{00}$ with respect to transverse momenta. The left and right panels correspond to $y_q=0$ and $y_q=1$, respectively. \label{fig:rho00_out_qxy}}
\end{figure}

On the left panel of Fig.~\ref{fig:rho00_out_qT_phi}, we plot $\rho^{\rm o}_{00}$ with $q_T$ dependence at fixed $y_q$ and $\phi$. Here $\rho^{\rm o}_{00}$ at $y_q=0$ and $y_q=1$ have similar behaviors with fixed $\phi$. In the case for $\phi=\pi/4$ ($q_x=q_y$), $\rho^{\rm o}_{00}<1/3$ and gradually approach $1/3$ at large $q_T$. However, with the maximum transverse momentum anisotropy at $\phi=0$ ($q_y=0$), it is found that $\rho^{\rm o}_{00}>1/3$ for sufficiently large $q_{T}$. Such an increase for $\rho^{\rm o}_{00}$ above $1/3$ could be  anticipated from Eq.~(\ref{eq:rho00o_large_gamma}) with large $\gamma$, which yields
\begin{eqnarray}\label{eq:rho00o_phi_large_gamma}
	\rho^{\rm o}_{00}(q)\approx \frac{1}{3}\bigg[1+\frac{2}{3}\gamma C_{BE2}\big(3\hat{v}_y^2+\hat{v}_z^2-1\big)\bigg]=\frac{1}{3}\bigg[1+\frac{2}{3}\gamma C_{BE2}\big(2\hat{v}_y^2-\hat{v}_x^2\big)\bigg]
\end{eqnarray} 
based on Eq.~(\ref{eq:CBE12_numerical}), where we have utilized $\hat{v}_x^2+\hat{v}_y^2+\hat{v}_z^2=1$ to acquire the second equality. It is clear to see that $\rho^{\rm o}_{00}>1/3$ when $\hat{v}_x^2>2\hat{v}_y^2$. On the contrary, when $\hat{v}_x^2< 2\hat{v}_y^2$ as the cases for $\phi=\pi/4$ and $\phi=0$, $\rho^{\rm o}_{00}<1/3$ at large momenta. Note $\gamma\approx 3.74\cosh y_q$ for $q_T=4$ GeV. On the right panel of Fig.~\ref{fig:rho00_out_qT_phi}, we plot  $\delta\rho^{\rm o}_{00}=\rho^{\rm o}_{00}-1/3$ with $\phi$ dependence at fixed $y_q$ and $q_T$.
Consistently, for larger $q_T$, one finds that $\delta\rho^{\rm o}_{00}$ at $\phi=0$ and $\phi=\pi$ (both for $q_y=0$) could be positive. In light of Eq.~(\ref{eq:rho00o_phi_large_gamma}), by taking $\hat{v}_x=\hat{v}_{T}\cos\phi$ and $\hat{v}_y=\hat{v}_{T}\sin\phi$ with $\hat{v}_T=\sqrt{\hat{v}_x^2+\hat{v}_y^2}$, it turns out that $\delta\rho^{\rm o}_{00}\sim -(2-3\cos^2\phi)$, which captures the qualitatively features of the results with large $q_T$.

In Fig.~\ref{fig:rho00_out_qT_yq}, we plot $\rho^{\rm o}_{00}$ with $y_q$ dependence at fixed $\phi$ and $q_T$. Generically, the $y_q$ dependence is relatively mild. Basically, $\rho^{\rm o}_{00}$ monotonically increases for $q_x=q_y$ and decreases for $q_y=0$. For large $q_T$, $\rho^{\rm o}_{00}$ could be larger than $1/3$ for $q_y=0$ as already indicated Fig.~\ref{fig:rho00_out_qT_phi}. We further show contour plots for $\delta\rho^{\rm o}_{00}$ with respect to transverse momenta in Fig.~\ref{fig:rho00_out_qxy}. In general, $\rho^{\rm o}_{00}<1/3$ in most of kinematic regions.

\begin{figure}[t]
	\includegraphics[scale=0.65]{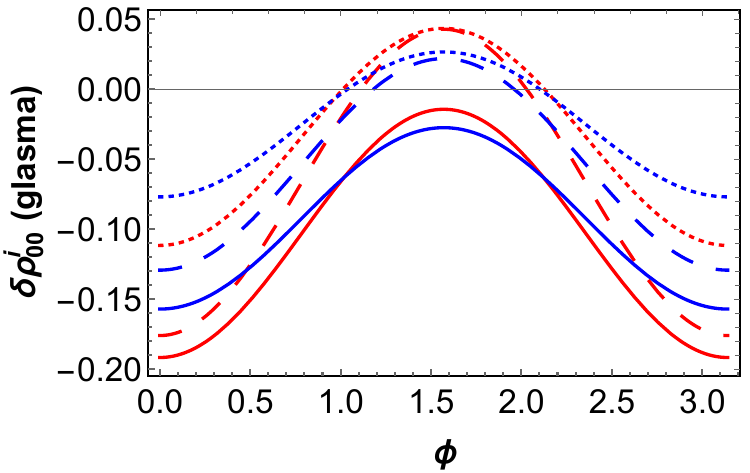}\quad\includegraphics[scale=0.65]{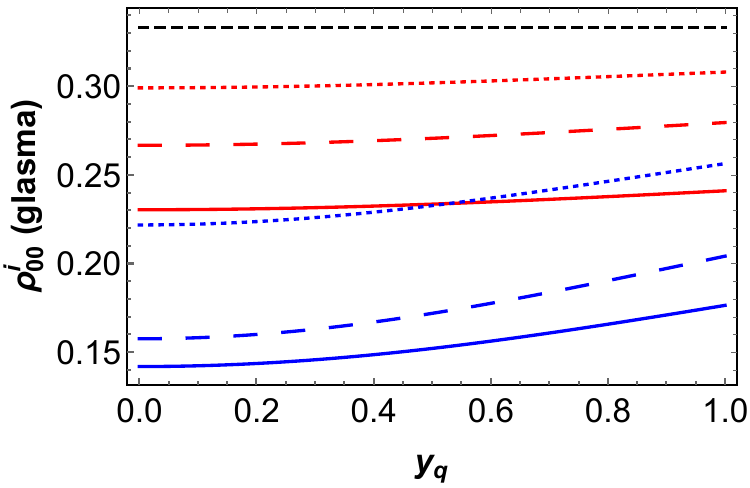}
	
	\caption{The left panel shows $\delta\rho^{\rm in}_{00}=\rho^{\rm o}_{00}-1/3$ with the azimuthal angle dependence with the same plot-style assignment in the right panel Fig.~\ref{fig:rho00_out_qT_phi}. The right panel shows $\rho^{\rm i}_{00}$ with $y_q$ dependence with the same plot-style assignment as Fig.~\ref{fig:rho00_out_qT_yq}.  \label{fig:rho00_in_phi_yq}}
\end{figure}
\begin{figure}[t]
	\includegraphics[scale=0.5]{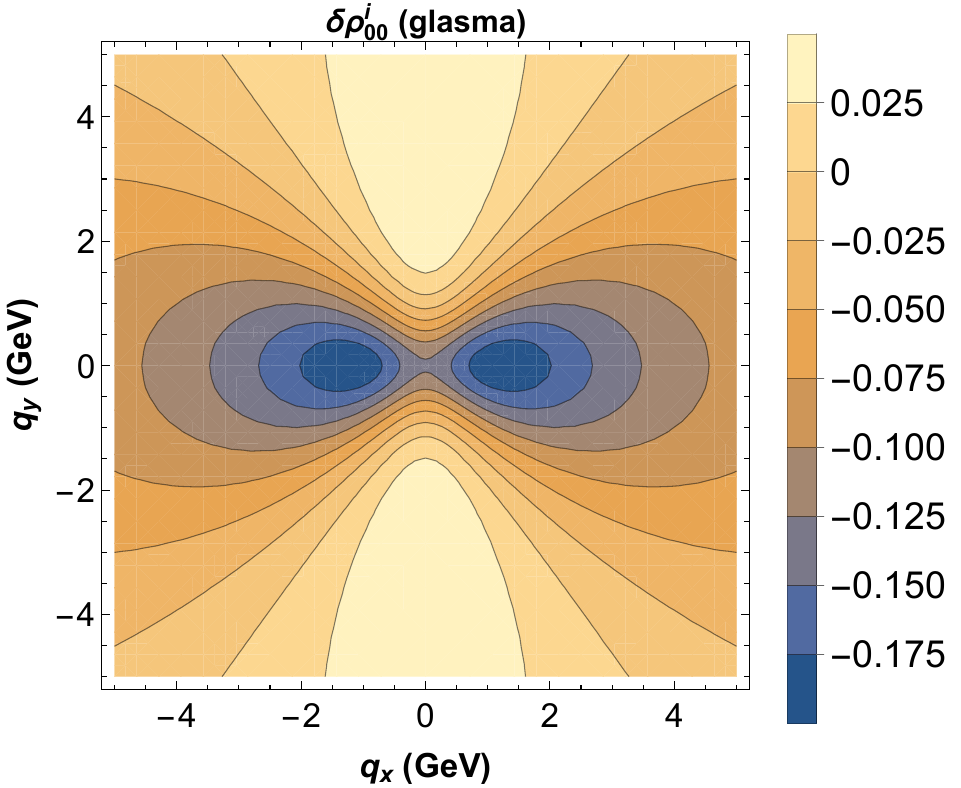}\quad\includegraphics[scale=0.5]{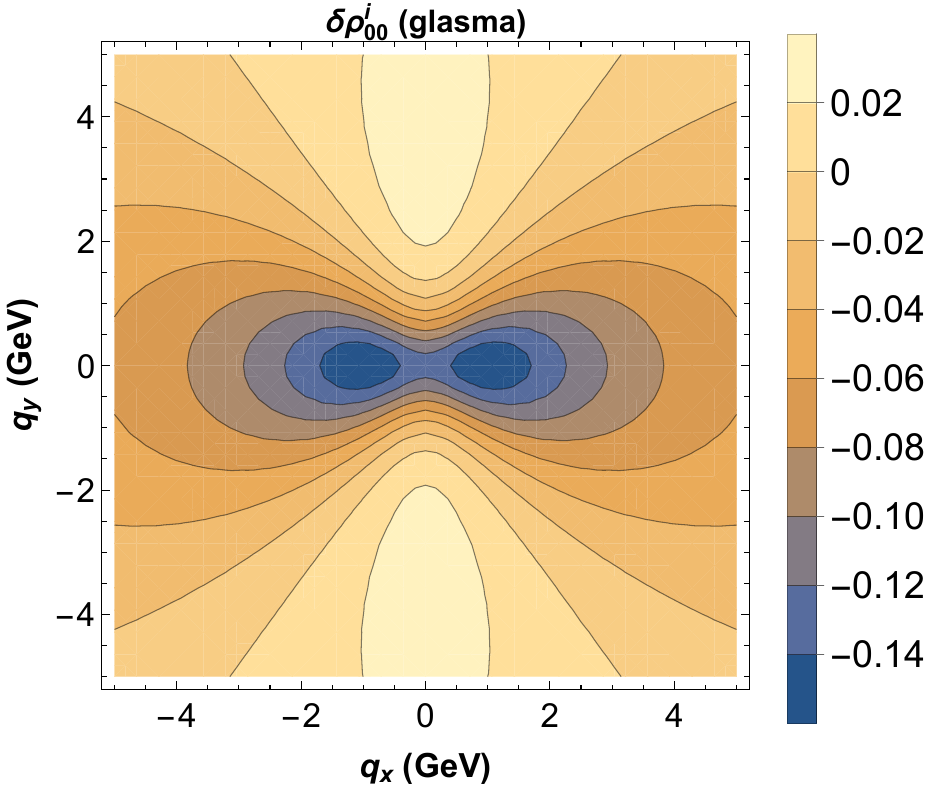}
	
	\caption{The contour plots for $\delta\rho^{\rm i}_{00}$ with respect to transverse momenta. The left and right panels correspond to $y_q=0$ and $y_q=1$, respectively. \label{fig:rho00_in_qxy}}
\end{figure}

Analogously, we may also numerically evaluate $\rho^{\rm i}_{00}$ for the in-plane spin alignment for $\phi$ mesons. Due to symmetry, the results for $\rho^{\rm i}_{00}$ with $q_T$ dependence at fixed $\phi$ and $y_q$ are the same as those shown in Fig.~\ref{fig:rho00_out_qT_phi} by exchanging the results for $\phi=0$ and  $\phi=\pi/2$. In Fig.~\ref{fig:rho00_in_phi_yq}, we further present $\delta\rho^{\rm i}_{00}$ with $\phi$ dependence and $\rho^{\rm i}_{00}$ with $y_q$ dependence, respectively. In particular, for the large-momentum limit, we simply need to exchange $\hat{v}_x$ and $\hat{v}_y$ in Eq.~(\ref{eq:rho00o_phi_large_gamma}), which yields $\delta\rho^{\rm o}_{00}\sim -(2-3\sin^2\phi)$, and the qualitatively features of $\delta\rho^{\rm i}_{00}$ with $\phi$ dependence at large $q_T$ are understood. Also, $\rho^{\rm i}_{00}<1/3$ for either $\phi=\pi$ or $\phi=0$ is expected. The $y_q$ dependence is generically mild. The contour plots in Fig.~\ref{fig:rho00_in_qxy} correspond to those in Fig.~\ref{fig:rho00_out_qxy} with $90$-degree rotations.

\begin{figure}[t]
	\includegraphics[scale=0.65]{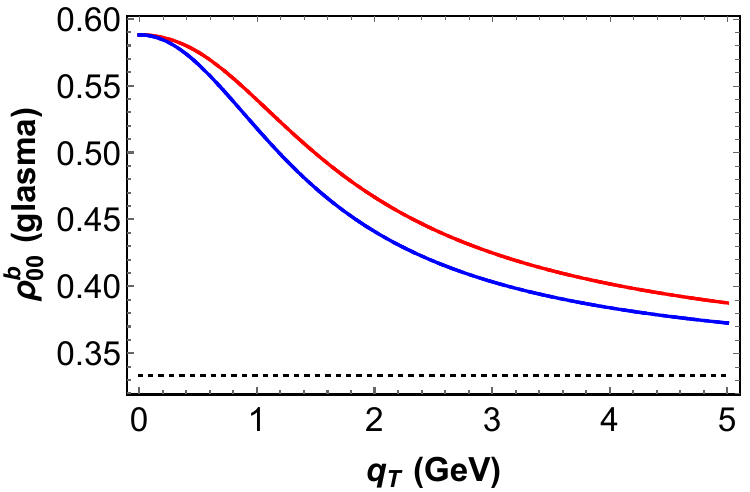}\quad\includegraphics[scale=0.65]{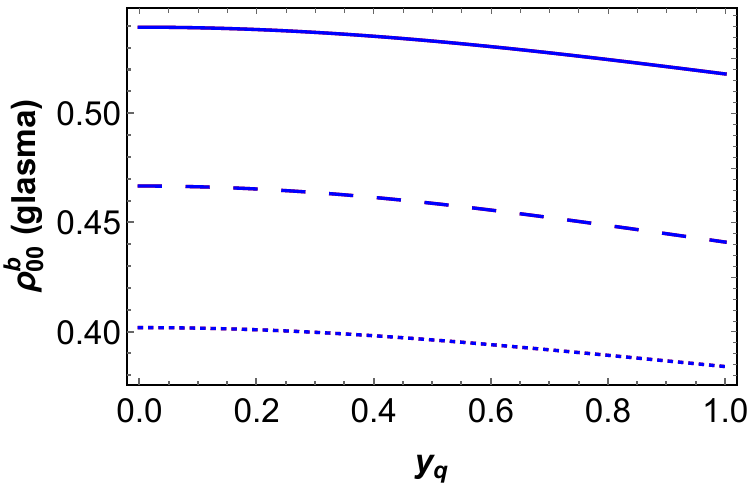}
	
	\caption{The left panel shows $\rho^{\rm b}_{00}$ with the transverse-momentum dependence and same plot-style assignment as the left panel of Fig.~\ref{fig:rho00_out_qT_phi}, while the curves from different $\phi$ coincide.  The right panel shows $\rho^{\rm b}_{00}$ with the momentum-rapidity dependence and same plot-style assignment as Fig.~\ref{fig:rho00_out_qT_yq}. Similarly, all curves with different $\phi$ collapse to one. \label{fig:rho00_beam_qT_yq}}
\end{figure}
\begin{figure}[t]
	\includegraphics[scale=0.5]{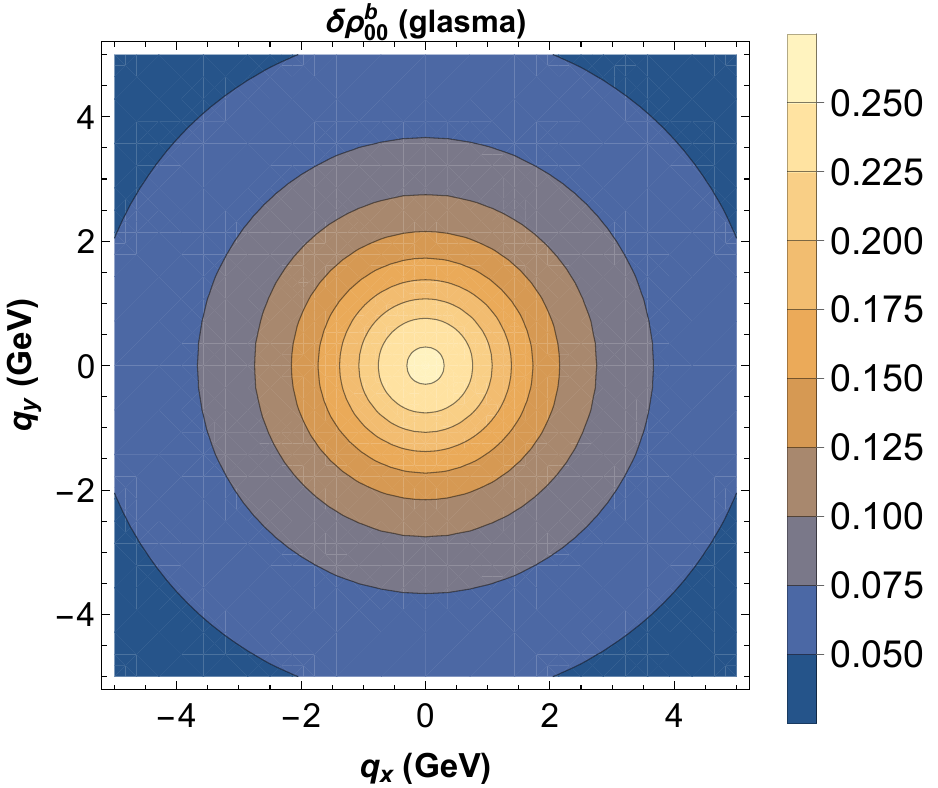}\quad\includegraphics[scale=0.5]{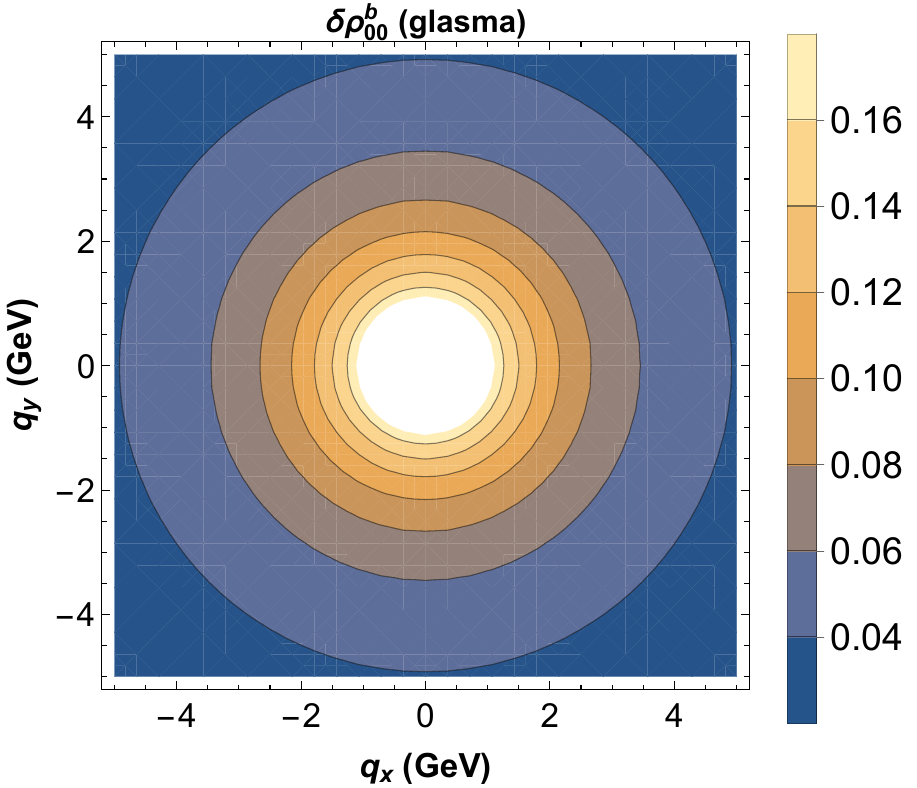}
	
	\caption{The contour plots for $\delta\rho^{\rm b}_{00}$ with respect to transverse momenta. The left and right panels correspond to $y_q=0$ and $y_q=1$, respectively. \label{fig:rho00_beam_qxy}}
\end{figure}
Finally, we present the results for choosing the beam direction as the spin quantization axis. In such a case, sharper differences with respect to $\rho^{\rm o}_{00}$ and $\rho^{\rm i}_{00}$ are found. As manifested by the degeneracy of $C^{\rm b}_x$ and $C^{\rm b}_y$ in Eq.~(\ref{eq:Cbxyz_coeff}), it is found that $\rho^{\rm b}_{00}$ is independent of $\phi$. In Fig.~\ref{fig:rho00_beam_qT_yq}, we present $\rho^{\rm b}_{00}$ with $q_T$ dependence and $y_q$ dependence, respectively. Now,  $\rho^{\rm b}_{00}$ is always greater than $1/3$ with mild $y_q$ dependence. As inferred from Eq.~(\ref{eq:rho00b_large_gamma}), $\rho^{\rm b}_{00}\rightarrow 1/3$ at large momenta given $|C_{BE2}|\gg |C_{BE1}|$. The contour plots of $\delta\rho^{\rm b}_{00}=\rho^{\rm b}_{00}-1/3$ are further shown in Fig.~\ref{fig:rho00_beam_qxy}, which have rather distinct features with respect to $\delta\rho^{\rm o}_{00}$ and $\delta\rho^{\rm i}_{00}$. Here $\delta\rho^{\rm b}_{00}$ respects the transversely rotational symmetry and becomes more prominent in the center with small momenta. 

\section{Spin alignment from color fields in QGP}\label{sec:spin_alignment_QGP}
In this section, we further investigate the spin alignment induced by chromo-electromagnetic fields in QGP or more precisely at the freeze-out hypersurface for chemical equilibrium in light of Eq.~(\ref{eq:Aa_QGP}). One distinct feature here is that such color fields are expected to be isotropic in the lab frame. 
When considering the spin alignment led by non-dynamical spin polarization without intrinsic anisotropy, Eq.~(\ref{eq:rho00_QKT_global}) for the out-of-plane spin alignment has to be slightly modified as 
\begin{eqnarray}
	\rho^{\rm o}_{00}(q)
	=\frac{1-\big[\big(\tilde{C}_{B1}+\tilde{C}_{B2}\big)\chi^{yy}_{B}+\tilde{C}_{B1}(\chi^{xx}_{B}+\chi^{zz}_{B}\big)+\big(\tilde{C}_{E1}+\tilde{C}_{E2}\big)\chi^{yy}_{E}+\tilde{C}_{E1}(\chi^{xx}_{E}+\chi^{zz}_{E}\big)\big]}{3-\big[\big(3\tilde{C}_{B1}+\tilde{C}_{B2}\big)\chi_{B}+\big(3\tilde{C}_{E1}+\tilde{C}_{E2}\big)\chi_{E}\big]},
\end{eqnarray}
where 
\begin{eqnarray}
\tilde{C}_{B1,2}=C_{B1,2}\frac{\hbar^2g^2C_{AB}}{\pi^2 m_qm_{\bar{q}}C_VN_c^2},\quad \tilde{C}_{E1,2}=C_{E1,2}\frac{\hbar^2g^2C_{AE}}{\pi^2 m_qm_{\bar{q}}C_VN_c^2},
\end{eqnarray}
and
\begin{eqnarray}
	C_{AB}&=&\frac{\big(\partial_{\epsilon_{q\tilde{k}}}f_{qV}(\epsilon_{q\tilde{k}},t_{\rm eq})\big)\big(\partial_{\epsilon_{\bar{q}\tilde{k}}}f_{\bar{q}V}(\epsilon_{\bar{q}\tilde{k}},t_{\rm eq})\big)}{16\epsilon_{q\tilde{k}}\epsilon_{\bar{q}\tilde{k}}},
	\\
	C_{AE}&=&\frac{1}{64\epsilon_{q\tilde{k}}\epsilon_{\bar{q}\tilde{k}}}
	\bigg(\big(\partial_{\epsilon_{q\tilde{k}}}f_{qV}(\epsilon_{q\tilde{k}},t_{\rm eq})\big)+\frac{f_{qV}(\epsilon_{q\tilde{k}},t_{\rm eq})}{\epsilon_{q\tilde{k}}}\bigg)\bigg(\big(\partial_{\epsilon_{\bar{q}\tilde{k}}}f_{\bar{q}V}(\epsilon_{\bar{q}\tilde{k}},t_{\rm eq})+\frac{f_{\bar{q}V}(\epsilon_{\bar{q}\tilde{k}},t_{\rm eq})}{\epsilon_{\bar{q}\tilde{k}}}\bigg).
\end{eqnarray}
Here the correlators of color fields are presumed to take the form 
\begin{eqnarray}
	\chi_B^{ij}= \big\langle B^{ai}(t_{\rm eq},\bm x)B^{aj}(t_{\rm eq},\bm x)\big\rangle\delta^{ij},\quad
	\chi_E^{ij}= \big\langle E^{ai}(t_{\rm eq},\bm x)E^{aj}(t_{\rm eq},\bm x)\big\rangle\delta^{ij},
\end{eqnarray}
where we consider equal-spacetime correlators or simply the constant color-field square at the freeze-out hypersurface for $x_0=t_{\rm eq}$. 
From Eqs.~(\ref{eq:BrBr_general}) and (\ref{eq:ErEr_general}), we have
\begin{eqnarray}\nonumber
\chi_B^{xx}(q)&=&\gamma^2\big(\chi_{0B}^{xx}+v_y^2\chi_{0E}^{zz}+v_z^2\chi_{0E}^{yy}\big)-2\gamma(\gamma-1)\chi_{0B}^{xx}\hat{v}_x^2+(\gamma-1)^2\hat{v}_x^2\chi_{0B},
\\\nonumber
\chi_B^{yy}(q)&=&\gamma^2\big(\chi_{0B}^{yy}+v_x^2\chi_{0E}^{zz}+v_z^2\chi_{0E}^{xx}\big)-2\gamma(\gamma-1)\chi_{0B}^{yy}\hat{v}_x^2+(\gamma-1)^2\hat{v}_y^2\chi_{0B},
\\
\chi_B^{zz}(q)&=&\gamma^2\big(\chi_{0B}^{zz}+v_y^2\chi_{0E}^{xx}+v_x^2\chi_{0E}^{yy}\big)-2\gamma(\gamma-1)\chi_{0B}^{zz}\hat{v}_z^2+(\gamma-1)^2\hat{v}_z^2\chi_{0B},
\end{eqnarray}
and
\begin{eqnarray}\nonumber
	\chi_E^{xx}(q)&=&\gamma^2\big(\chi_{0E}^{xx}+v_y^2\chi_{0B}^{zz}+v_z^2\chi_{0B}^{yy}\big)-2\gamma(\gamma-1)\chi_{0E}^{xx}\hat{v}_x^2+(\gamma-1)^2\hat{v}_x^2\chi_{0E},
	\\\nonumber
	\chi_E^{yy}(q)&=&\gamma^2\big(\chi_{0E}^{yy}+v_x^2\chi_{0B}^{zz}+v_z^2\chi_{0B}^{xx}\big)-2\gamma(\gamma-1)\chi_{0E}^{yy}\hat{v}_x^2+(\gamma-1)^2\hat{v}_y^2\chi_{0E},
	\\
	\chi_E^{zz}(q)&=&\gamma^2\big(\chi_{0E}^{zz}+v_y^2\chi_{0B}^{xx}+v_x^2\chi_{0B}^{yy}\big)-2\gamma(\gamma-1)\chi_{0E}^{zz}\hat{v}_z^2+(\gamma-1)^2\hat{v}_z^2\chi_{0E},
\end{eqnarray}
where $\chi_{0B,E}^{ij}=\chi_{B,E}^{ij}(q=0)$ and  $\chi_{0B,E}=\chi_{0B,E}^{xx}+\chi_{0B,E}^{yy}+\chi_{0B,E}^{zz}$. 
When further postulating  $\chi_{0B}^{ii}=\chi_{0E}^{ii}=\chi_{0}$ such that the isotropic chromo-electric correlator and the chromo-magnetic one are further degenerate, it is found
\begin{eqnarray}\nonumber
	\chi_B^{xx}(q)&=&\chi_E^{xx}(q)=\big(2\gamma^2-1+4(1-\gamma)\hat{v}_x^2)\big)\chi_{0},
	\\\nonumber
	\chi_B^{yy}(q)&=&\chi_E^{yy}(q)=\big(2\gamma^2-1+4(1-\gamma)\hat{v}_y^2)\big)\chi_{0},
	\\
	\chi_B^{zz}(q)&=&\chi_E^{zz}(q)=\big(2\gamma^2-1+4(1-\gamma)\hat{v}_z^2)\big)\chi_{0}.
\end{eqnarray}
For the out-of-plane spin alignment, it turns out that 
\begin{eqnarray}\nonumber\label{eq:rho00_out_QGP}
	\rho^{\rm o}_{00}(q)
	&=&\frac{1-\big[\big(\tilde{C}_{BE1}+\tilde{C}_{BE2}\big)\chi^{yy}_{B}+\tilde{C}_{BE1}(\chi^{xx}_{B}+\chi^{zz}_{B}\big)\big]}{3-\big(3\tilde{C}_{BE1}+\tilde{C}_{BE2}\big)\chi_{B}}
	\\
	&=&\frac{1-\big(3\tilde{C}_{BE1}+\tilde{C}_{BE2}\big)(2\gamma^2-1)-4(1-\gamma)(\tilde{C}_{BE1}+\tilde{C}_{BE2}\hat{v}_y^2)}{3-\big(3\tilde{C}_{BE1}+\tilde{C}_{BE2}\big)(6\gamma^2-4\gamma+1)}
	,
\end{eqnarray}
where $\tilde{C}_{BE1,2}=(\tilde{C}_{B1,2}+\tilde{C}_{E1,2})\chi_{0}$. Similarly, it is found
\begin{eqnarray}
	\rho^{\rm i}_{00}(q)
	&=&\frac{1-\big(3\tilde{C}_{BE1}+\tilde{C}_{BE2}\big)(2\gamma^2-1)-4(1-\gamma)(\tilde{C}_{BE1}+\tilde{C}_{BE2}\hat{v}_x^2)}{3-\big(3\tilde{C}_{BE1}+\tilde{C}_{BE2}\big)(6\gamma^2-4\gamma+1)}
\end{eqnarray}
for the out-of-plane spin alignment and
\begin{eqnarray}
	\rho^{\rm b}_{00}(q)
	&=&\frac{1-\big(3\tilde{C}_{BE1}+\tilde{C}_{BE2}\big)(2\gamma^2-1)-4(1-\gamma)(\tilde{C}_{BE1}+\tilde{C}_{BE2}\hat{v}_z^2)}{3-\big(3\tilde{C}_{BE1}+\tilde{C}_{BE2}\big)(6\gamma^2-4\gamma+1)}
\end{eqnarray}
for the longitudinal spin alignment.

\begin{figure}[t]
	\includegraphics[scale=0.65]{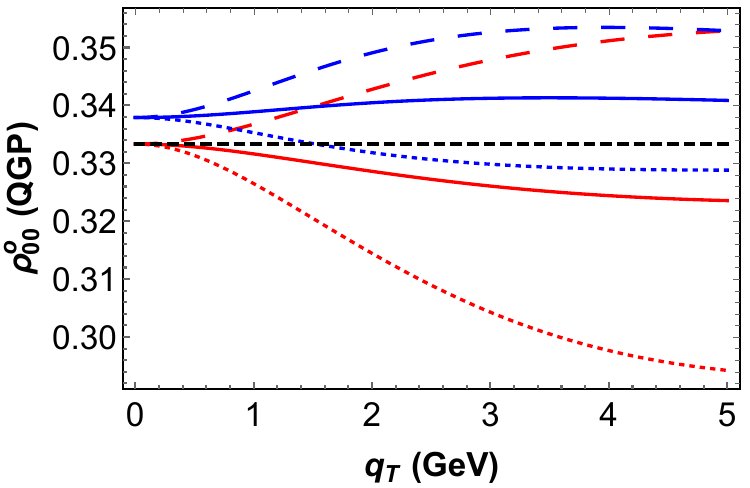}\quad\includegraphics[scale=0.65]{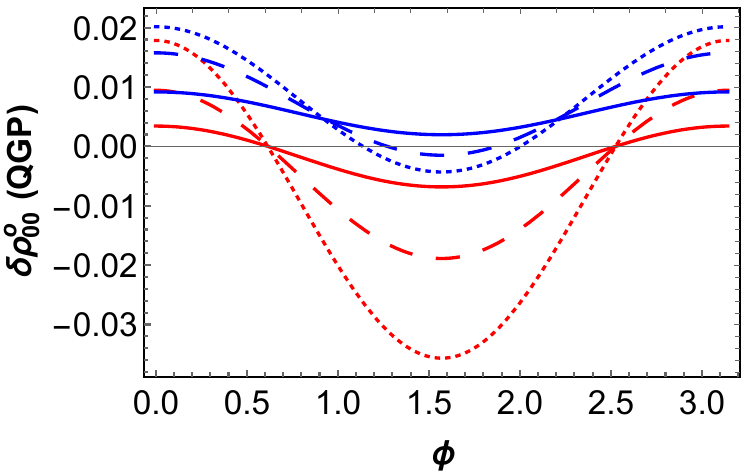}
	
	\caption{Same plot setting as Fig.~\ref{fig:rho00_out_qT_phi} for the out-of-plane spin alignment from QGP. \label{fig:QGP_rho00_out_qT_phi}}
\end{figure}
\begin{figure}[t]
	\includegraphics[scale=0.65]{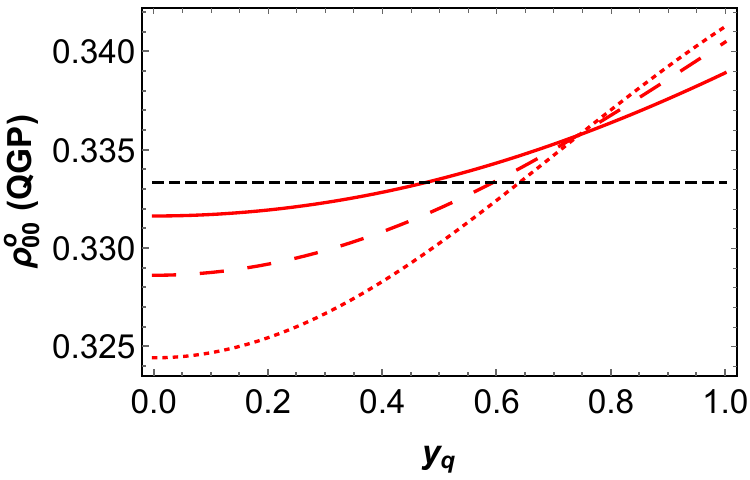}\quad\includegraphics[scale=0.65]{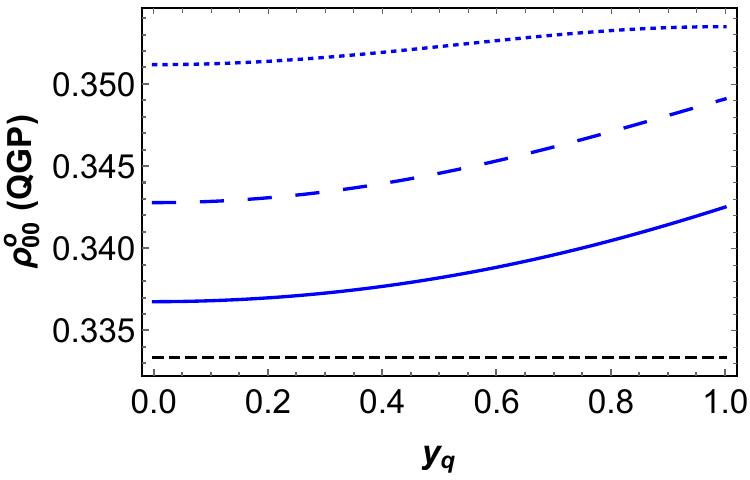}
	
	\caption{Same plot setting as Fig.~\ref{fig:rho00_out_qT_yq} for the out-of-plane spin alignment from QGP. \label{fig:QGP_rho00_out_qT_yq}}
\end{figure}
\begin{figure}[t]
	\includegraphics[scale=0.5]{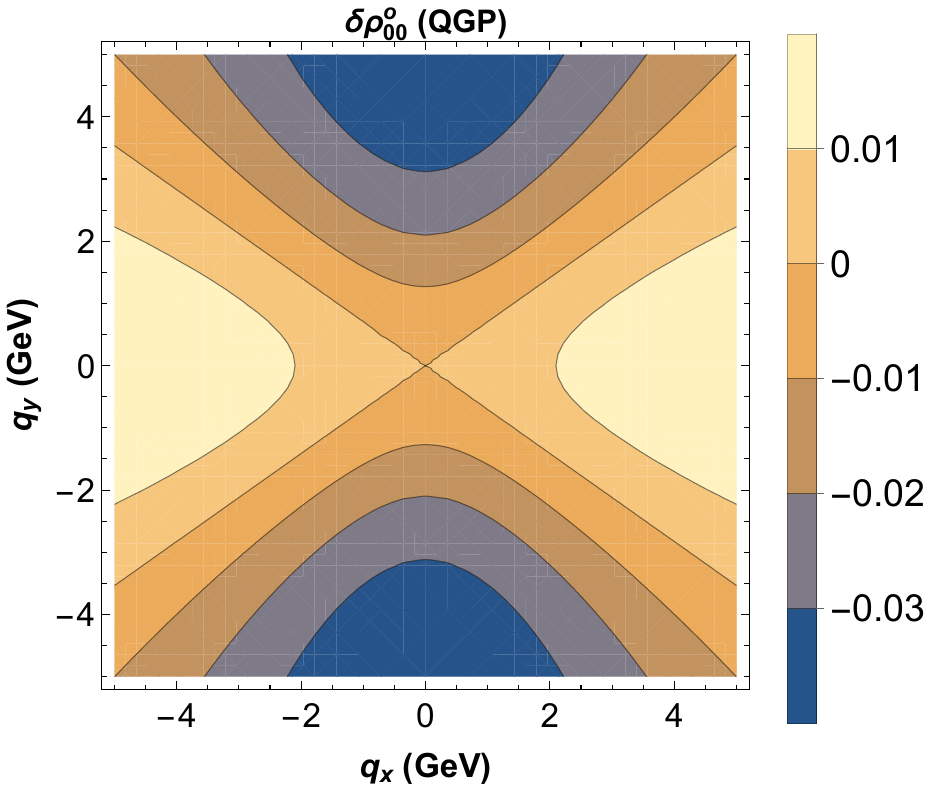}\quad\includegraphics[scale=0.5]{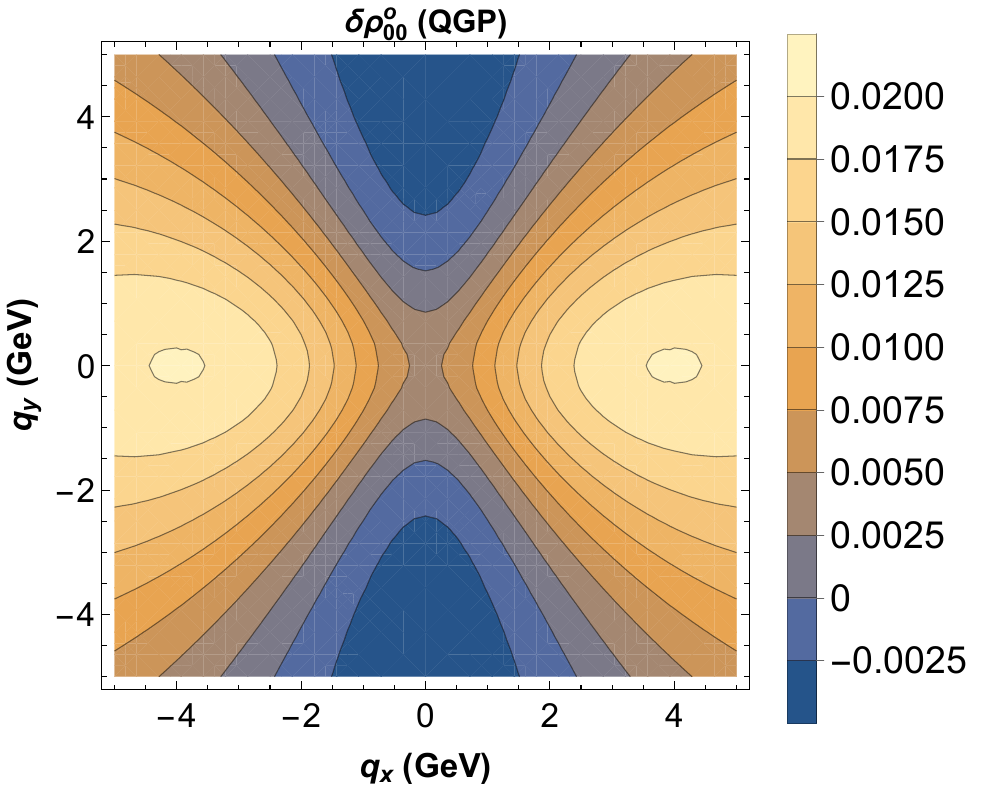}
	
	\caption{Same plot setting as Fig.~\ref{fig:rho00_out_qxy} for the out-of-plane spin alignment from QGP. \label{fig:QGP_rho00_out_qxy}}
\end{figure}
Subsequently, we may adopt the same numerical parameters as in the previous section to evaluate spin alignment with different spin quantization axes. However, unlike the case for glasma fields, here we simply approximate $\chi_0=g^2T_{\rm eq}^4$ as the strength of color-field correlators in QGP at the freeze-out hypersurface for simplicity and take $\alpha_s=g^2/(4\pi)=1/3$. Recall that the primary purpose in this section is to analyze the qualitative features instead of making precise predictions quantitatively. In Fig.~\ref{fig:QGP_rho00_out_qT_phi}, we plot $\rho^{\rm o}_{00}$ with $q_T$ dependence and $\delta \rho^{o}_{00}$ with $\phi$ dependence as a comparison with those from glasma fields in Fig.~\ref{fig:rho00_out_qT_phi}. Unlike the results from glasma, now the spin alignment is induced by finite momenta as  observed from the red curves in the left panel of Fig.~\ref{fig:QGP_rho00_out_qT_phi}, where $\rho^{\rm o}_{00}=1/3$ when $q_T=0$ and $y_q=0$. Given Eq.~(\ref{eq:CBE12_numerical}) and correspondingly $\tilde{C}_{BE2}<0$ and $|\tilde{C}_{BE2}|\gg \tilde{C}_{BE1}>0$ for $\phi$ mesons, we may approximate Eq.~(\ref{eq:rho00_out_QGP}) as
\begin{eqnarray}
	\rho^{\rm o}_{00}(q)
	&\approx &\frac{1-\tilde{C}_{BE2}\big(2\gamma^2-1+4(1-\gamma)\hat{v}_y^2\big)}{3-\tilde{C}_{BE2}(6\gamma^2-4\gamma+1)}
	,
\end{eqnarray}
which further reduces to 
\begin{eqnarray}\label{eq:rho00_out_QGP_smallC}
\rho^{\rm o}_{00}(q)\approx \frac{1}{3}\Big[1+4\tilde{C}_{BE2}(\gamma-1)\Big(\hat{v}_y^2-\frac{1}{3}\Big)\Big]
\end{eqnarray}
for weak correlations. According to Eq.~(\ref{eq:rho00_out_QGP_smallC}), one immediately finds $\rho^{\rm o}_{00}>1/3$ when $\hat{v}_y^2<1/3$ especially at the large-momentum limit. The $\phi$ dependence of $\rho^{\rm o}_{00}$ is then shown on the right panel of Fig.~\ref{fig:QGP_rho00_out_qT_phi}. Nevertheless, from Eq.~(\ref{eq:rho00_out_QGP_smallC}) with larger $\gamma$, we also obtains $\delta\rho^{\rm o}_{00}\sim -(2-3\cos^2\phi)$ as the same qualitative feature in the case with glasma fields. We also plot $\rho^{\rm o}_{00}$ with the momentum-rapidity dependence in Fig.~\ref{fig:QGP_rho00_out_qT_yq}. One observes the monotonic increase with respect to $y_q$ especially for $\phi=\pi/4$ shown on the left panel. Finally, the contour plots with transverse-momentum dependence are shown in Fig.~\ref{fig:QGP_rho00_out_qxy}, where now $\delta\rho^{\rm o}_{00}$ is more prominent away from the center with small momenta, which is also anticipated since the spin alignment is now triggered by momentum anisotropy. Overall, the out-of-plane spin alignment from the isotropic color fields in QGP is qualitatively consistent with the one from fluctuating meson fields in Refs.~\cite{Sheng:2022wsy,Sheng:2023urn}.
When comparing the results here with those from the glasma effect, one may particularly focus on the $q_T$ dependence and also the detailed transverse-momenta dependence shown in the contour plots for $\rho^{\rm o}_{00}$, where the distinct qualitative features are found from our analyses.  
\begin{figure}[t]
	\includegraphics[scale=0.65]{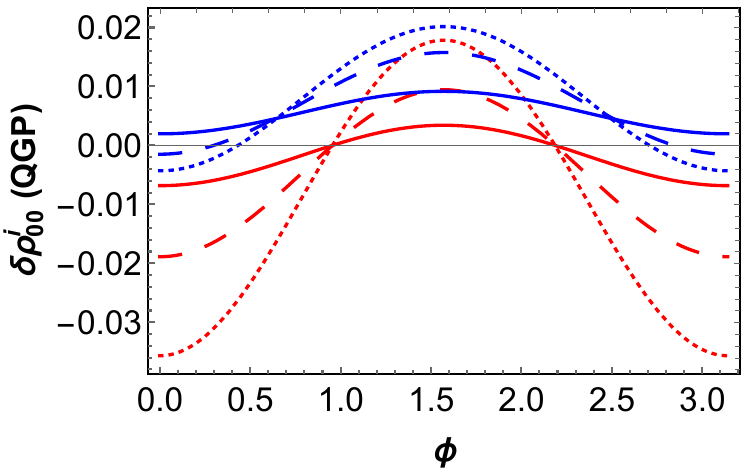}\quad\includegraphics[scale=0.65]{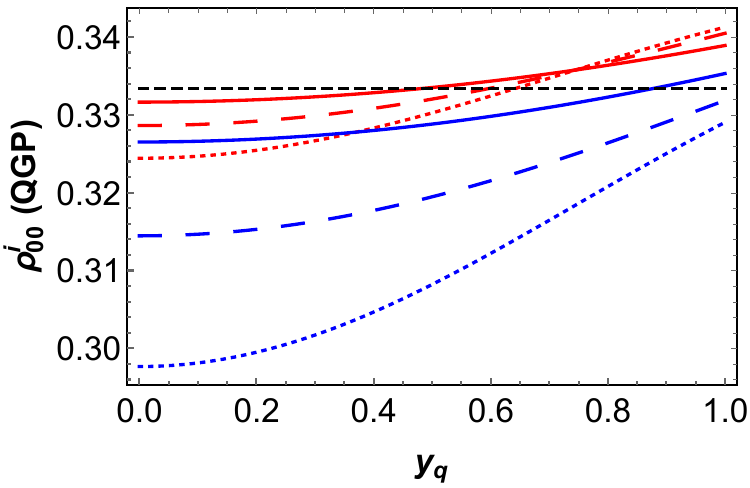}
	
	\caption{Same plot setting as Fig.~\ref{fig:rho00_in_phi_yq} for the in-plane spin alignment from QGP. \label{fig:QGP_rho00_in_phi_yq}}
\end{figure}
\begin{figure}[t]
	\includegraphics[scale=0.5]{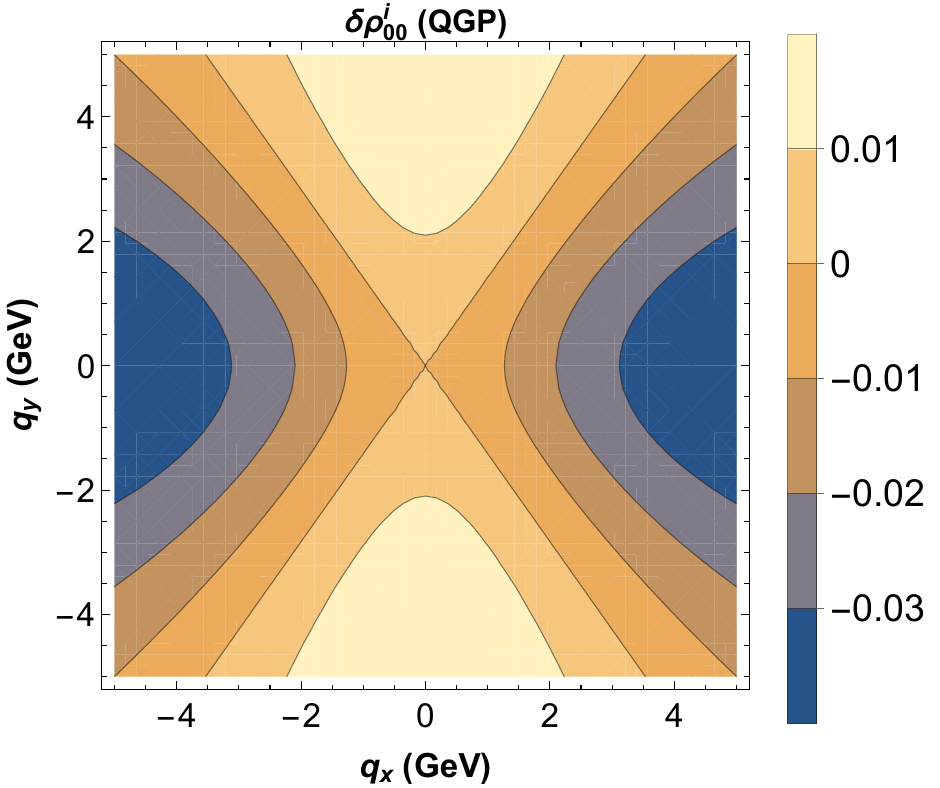}\quad\includegraphics[scale=0.5]{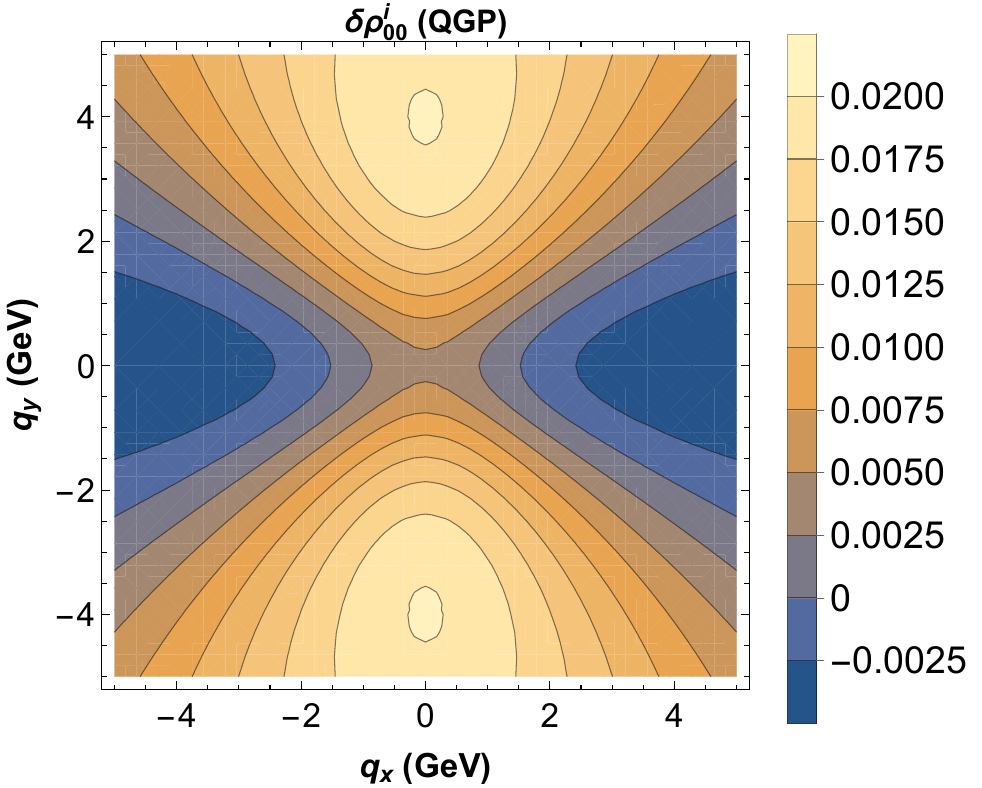}
	
	\caption{Same plot setting as Fig.~\ref{fig:rho00_in_qxy} for the in-plane spin alignment from QGP. \label{fig:QGP_rho00_in_qxy}}
\end{figure}

\begin{figure}[t]
	\includegraphics[scale=0.65]{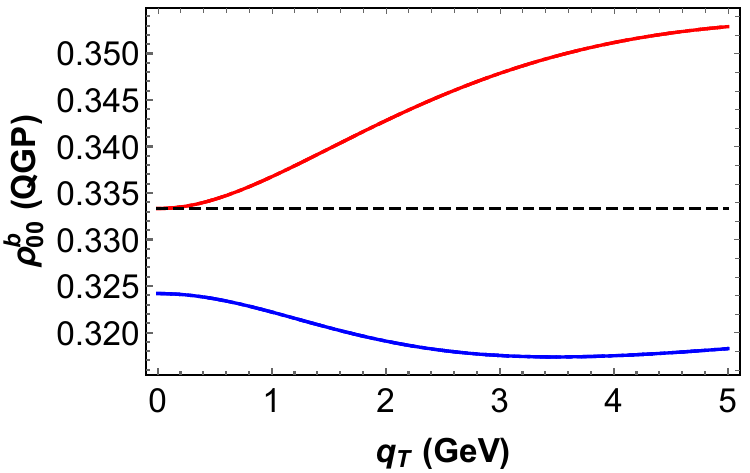}\quad\includegraphics[scale=0.65]{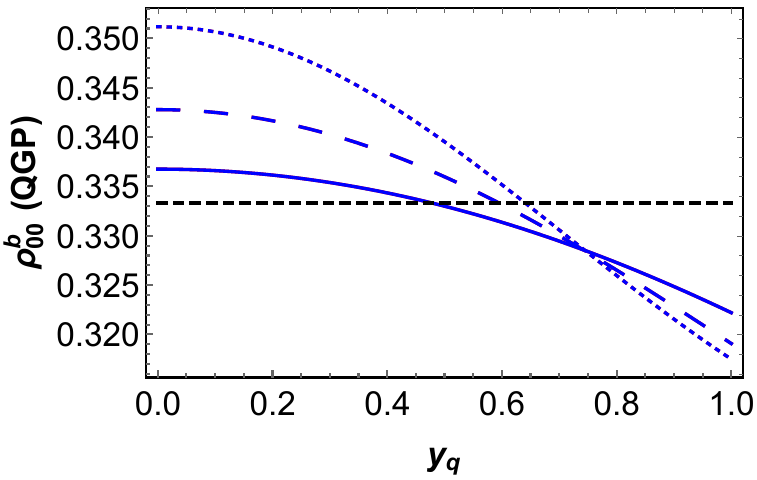}
	
	\caption{Same plot setting as Fig.~\ref{fig:rho00_beam_qT_yq} for the longitudinal spin alignment from QGP.  \label{fig:QGP_rho00_beam_qT_yq}}
\end{figure}
\begin{figure}[t]
	\includegraphics[scale=0.5]{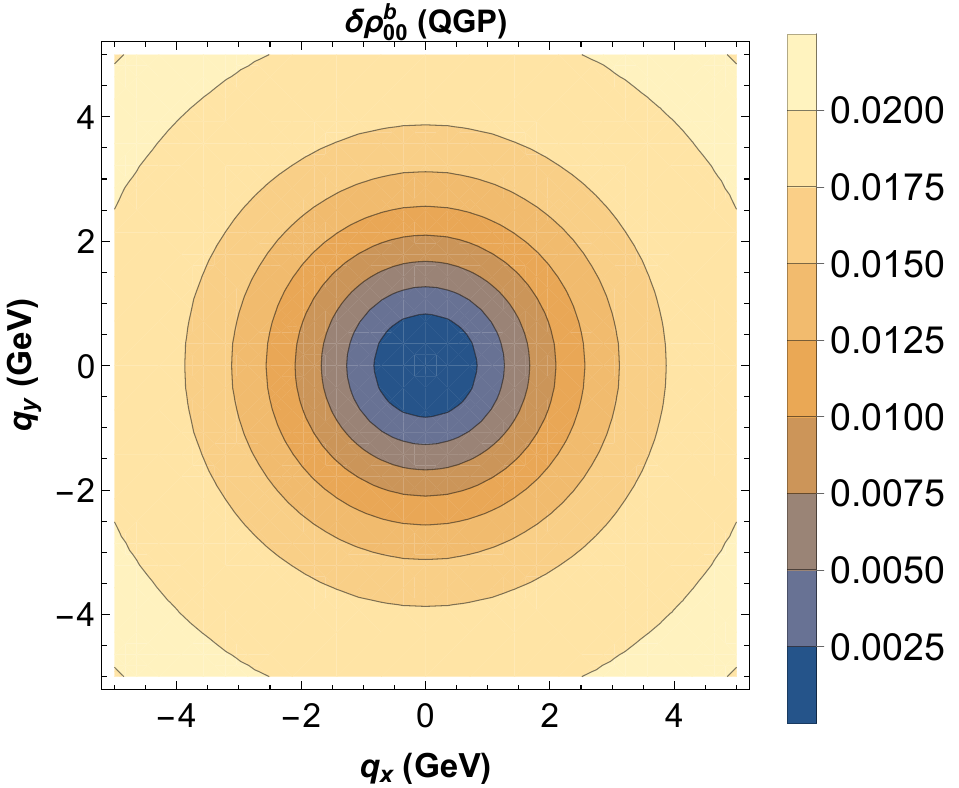}\quad\includegraphics[scale=0.5]{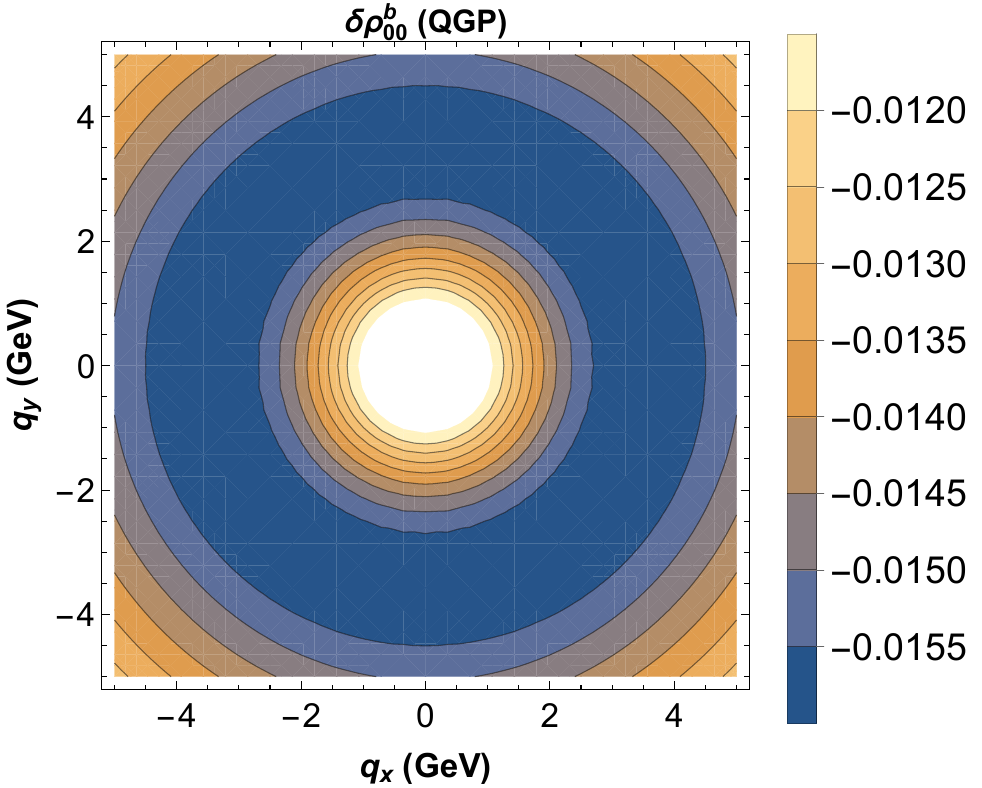}
	
	\caption{Same plot setting as Fig.~\ref{fig:QGP_rho00_beam_qxy} for the longitudinal spin alignment from QGP.  \label{fig:QGP_rho00_beam_qxy}}
\end{figure}
One may analogously evaluate $\rho^{\rm i}_{00}$, for which we present some of numerical results in Fig.~\ref{fig:QGP_rho00_in_phi_yq} and Fig.~\ref{fig:QGP_rho00_in_qxy}. Several features can be explained by symmetry as the case for the glasma effect. Finally, we evaluate the longitudinal spin alignment for isotropic color fields from QGP, for which there exists no $\phi$ dependence by symmetry. On the left panel of Fig.~\ref{fig:QGP_rho00_beam_qT_yq}, the $q_T$ dependence of $\rho^{\rm b}_{00}$ is shown, for which the qualitative behaviors may be captured by the approximated form in Eq.~(\ref{eq:rho00_out_QGP_smallC}) by replacing $\hat{v}_y$ with $\hat{v}_z$. Since $\hat{v}_z=0$ for $y_q=0$, it is clear to see that $\delta\rho^{\rm b}_{00}$ monotonically increases from zero stemming from the $\gamma-1$ prefactor. For $y_q=1$ and $\hat{v}_z^2>1/3$, $\delta\rho^{\rm b}_{00}$ is negative at $q_T=0$, while the variation with $q_T$ is more subtle since $\hat{v}_z$ also depends on $q_T$. The $y_q$ dependence at fixed $q_T$ is also shown on the right panel of Fig.~\ref{fig:QGP_rho00_beam_qT_yq}. Here $\delta\rho^{\rm b}_{00}$ can change from positive to negative when $y_q$ changes from $\hat{v}_z^2<1/3$ to $\hat{v}_z^2>1/3$. Eventually, the contour plots with transverse-momenta dependence are shown in Fig.~\ref{fig:QGP_rho00_beam_qxy}. As already discussed, the sign of $\delta\rho^{\rm b}_{00}$ depends on $y_q$, whereas  $\delta\rho^{\rm b}_{00}$ is more prominent away from the center with small momenta as a generic feature from isotropic color fields as opposed to the case for anisotropic color fields from the glasma. It is evident that the longitudinal spin alignment shows drastically qualitative difference between the color fields from QGP and glasma.   

\section{Conclusions and outlook}\label{sec:final}
In this paper, we study the momentum dependence of transverse and longitudinal spin alignment for $\phi$ mesons induced by the color fields from both the glasma phase and QGP in the quark coalescence scenarios. For the spin alignment from the glasma effect, it is found that large transverse-momentum anisotropy could overcome the effect from intrinsic anisotropy led by the longitudinally dominant glasma fields at large transverse momenta, which results in the sign flipping of $\delta \rho_{00}$ as the deviation from the baseline $\rho_{00}=1/3$ at intermediate momenta for transverse spin alignment and especially the out-of-plane case. Moreover, for longitudinal spin alignment from glasma fields, the azimuthal-angle dependence vanishes and its transverse-momentum dependence has rather distinct patterns with respect to the cases for transverse spin alignment. For the transverse spin alignment induced by isotropic color fields in the QGP, we find similar patterns qualitatively consistent with that by fluctuating vector-meson fields in the previous studies, while several differences from the glasma effect are more manifested with momentum dependence. On the other hand, the qualitative differences between the glasma effect and the color fields in QGP are even more robust for longitudinal spin alignment. In the following, we further discuss some potential issues about our approach and future directions.      

For more straightforward comparisons with present and future experimental observations, $\rho_{00}$ weighted by the particle momentum spectra is needed. But qualitatively, the azimuthal-angle dependence for $0\leq \phi\leq \pi/4$ should be correlated with the elliptic flow or indirectly the centrality dependence from more peripheral $(\phi\rightarrow 0)$ to more central $(\phi\rightarrow \pi/4)$ collisions. Also, when integrating over $q_T$, one may expect more dominant contributions from the low-$q_T$ region due to the weighting. Nevertheless, before conducting such more refined studies in phenomenology, we may have to first address several fundamental limitations and insufficiency regarding our theoretical approach as will be partly elaborated below.   
  
Regarding the flavor dependence, although we have lifted the equal-mass condition for the collision kernel responsible for the quark coalescence, we have not applied the formalism to vector mesons composed of quarks and antiquarks with different flavors. For $K^{*0}$, due to the short lifetime, they may further decay and regenerate in the hadronic phase \cite{Shapoval:2017jej,ALICE:2019xyr}, for which the spin correlation built up in even the QGP phase could be modified. On the other hand, it may be questionable for the validity of QKT for light quarks with quasi-particle approximation in the strongly coupled QGP. Assuming the spin correlation can be induced by glasma fields in the pre-equilibrium phase, it is curious and indispensable to further investigate how such a microscopic quantity could be incorporated in a dissipative effective theory such as the viscous hydrodynamics to address the spin relaxation in a non-perturbative manner in QGP. 

In fact, the weak-field approximation adopted to perturbatively solve the QKT could also be invalid given the glasma fields are actually non-perturbative. For example, a more thorough analysis for the competition between the color-octet and color-singlet contributions has been previously presented in Ref.~\cite{Kumar:2023ghs}. Generically, the QKT should be more applicable to heavy flavor, for which the scale separation is achievable. The suitable candidates such as the spin alignment of $J/\psi$ has been recently measured \cite{ALICE:2022dyy}. However, the quarkonia are more deeply bounded than light mesons. They undergo the screening, dynamical dissociation, and even recombination in the QGP. A recent study upon the construction of polarization-dependent QKT for vector quarkonia from open quantum systems and effective field theories was done in Ref.~\cite{Yang:2024ejk}, from which one may also investigate the spin alignment triggered by color fields in the QGP phase in a non-perturbative fashion. Nevertheless, how to incorporate the early-time spin correlation before the QGP phase in such a framework is unknown and the glasma effect for spin alignment also highly depends on the initial quark and antiquark distributions. In conclusion, tremendous efforts are still needed to understand the spin alignment of vector mesons with the underlying mechanisms in connection to the QCD interaction and microscopic properties of QGP or even the pre-equilibrium phase in high-energy nuclear collisions.              

\acknowledgments
The author would like to thank the hospitality of ECT* during the workshop “Spin and Quantum Features of QCD Plasma” and other participants, especially A. Tang, for helpful discussions.
This work is supported by National Science and Technology Council (Taiwan) under Grants No. MOST 110-2112-M-001-070-MY3 and No. NSTC 113-2628-M-001-009-MY4 and by Academia Sinica under Project No. AS-CDA-114-M01. 

\appendix
\section{Detailed calculations for quark coalescence}\label{app:quark_coalescence}
Eqs.~(\ref{eq:useful_rel_1}) and (\ref{eq:useful_rel_2}) yield
\begin{eqnarray}\label{eq:rel_1}
	-\frac{2|k\cdot\epsilon(\lambda, {\bm q})|^2}{N_m}\rightarrow -\frac{2|\bm k|^2\bar{z}^2}{N_m},
\end{eqnarray} 
\begin{eqnarray}\nonumber\label{eq:2ndrel_1}
	\mathcal{A}_{q}(\bm p,x)\cdot \mathcal{A}_{\bar{q}}(\bm p',x)&\rightarrow& \eta^{ij}\mathcal{A}_{qi}^{(0)}\mathcal{A}_{\bar{q}j}^{(0)}
	-\frac{|\bm k|^2}{\epsilon_{q\bm k}\epsilon_{\bar{q}\bm k}}\Big[\bar{z}^2\epsilon^{i}_{\lambda}\epsilon^{j}_{\lambda}-\frac{(1-\bar{z}^2)}{2}\hat{\Theta}^{ij}_{\lambda}\Big]\mathcal{A}_{qi}^{(0)} \mathcal{A}_{\bar{q}j}^{(0)}
	\\\nonumber
	&&-|\bm k|^2\Big[\bar{z}^2\epsilon^{i}_{\lambda}\epsilon^{j}_{\lambda}-\frac{(1-\bar{z}^2)}{2}\hat{\Theta}^{ij}_{\lambda}\Big]\mathcal{A}_{qil}^{(1)}\mathcal{A}_{\bar{q}j}^{(1)l}
	\\\nonumber
	&=&\eta^{ij}\mathcal{A}_{qi}^{(0)} \mathcal{A}_{\bar{q}j}^{(0)}\left[1+\frac{|\bm k|^2(1-\bar{z}^2)}{2\epsilon_{q\bm k}\epsilon_{\bar{q}\bm k}}\right]
	+\frac{|\bm k|^2(1-3\bar{z}^2)}{2\epsilon_{q\bm k}\epsilon_{\bar{q}\bm k}}\epsilon^{i}_{\lambda}\epsilon^{j}_{\lambda}\mathcal{A}_{qi}^{(0)} \mathcal{A}_{\bar{q}j}^{(0)}
	\\
	&&+\frac{|\bm k|^2}{2}\big[(1-\bar{z}^2)\mathcal{A}_{qij}^{(1)}\mathcal{A}_{\bar{q}}^{(1)ij}+(1-3\bar{z}^2)\epsilon^{i}_{\lambda}\epsilon^{j}_{\lambda}\mathcal{A}_{qil}^{(1)}\mathcal{A}_{\bar{q}j}^{(1)l}\big]
	,
\end{eqnarray}
\begin{eqnarray}\nonumber
	&&-\frac{2|k\cdot\epsilon(\lambda, {\bm q})|^2}{N_m}\mathcal{A}_{q}(\bm p,x)\cdot \mathcal{A}_{\bar{q}}(\bm p',x)
	\\\nonumber
	&&\rightarrow -\frac{2|\bm k|^2\bar{z}^2}{N_m}\eta^{ij}\mathcal{A}_{qi}^{(0)}\mathcal{A}_{\bar{q}j}^{(0)}
	+\frac{2|\bm k|^4}{N_m\epsilon_{q\bm k}\epsilon_{\bar{q}\bm k}}\hat{\Theta}^{ijln}_{\lambda}\epsilon_{\lambda i}\epsilon_{\lambda j}\mathcal{A}_{ql}^{(0)}\mathcal{A}_{qn}^{(0)}
	+\frac{2|\bm k|^4}{N_m}\hat{\Theta}^{ijln}_{\lambda}\epsilon_{\lambda i}\epsilon_{\lambda j}\mathcal{A}_{qlk}^{(1)}\mathcal{A}_{\bar{q}n}^{(1)k}
	\\\nonumber
	&&=-\frac{2|\bm k|^2}{N_m}\bigg[\bar{z}^2\eta^{ij}\mathcal{A}_{qi}^{(0)} \mathcal{A}_{\bar{q}j}^{(0)}\left(1+\frac{|\bm k|^2(1-\bar{z}^2)}{2\epsilon_{q\bm k}\epsilon_{\bar{q}\bm k}}\right)
	-\frac{|\bm k|^2\bar{z}^2(1-7\bar{z}^2)}{2\epsilon_{q\bm k}\epsilon_{\bar{q}\bm k}}\epsilon^{i}_{\lambda}\epsilon^{j}_{\lambda}\mathcal{A}_{qi}^{(0)} \mathcal{A}_{\bar{q}j}^{(0)}
	\\
	&&\quad +\frac{|\bm k|^2\bar{z}^2}{2}
	\big((1-\bar{z}^2)\mathcal{A}_{qij}^{(1)}\mathcal{A}_{\bar{q}}^{(1)ij}+(1-7\bar{z}^2)\epsilon^{i}_{\lambda}\epsilon^{j}_{\lambda}\mathcal{A}_{qil}^{(1)}\mathcal{A}_{\bar{q}j}^{(1)l}\big)
	\bigg],
\end{eqnarray}
\begin{eqnarray}\nonumber
	&&2{\rm Re}\Big(\epsilon(\lambda,{\bm q})\cdot\mathcal{A}_{q}(\bm p,x)\epsilon^*(\lambda,{\bm q})\cdot\mathcal{A}_{\bar{q}}(\bm p',x)\Big)
	\\\nonumber
	&&\rightarrow 2\epsilon_{\lambda}\cdot\mathcal{A}_{q}^{(0)}\epsilon_{\lambda}\cdot \mathcal{A}_{\bar{q}}^{(0)}
	-2|\bm k|^2\Big[\bar{z}^2\epsilon^{i}_{\lambda}\epsilon^{j}_{\lambda}-\frac{(1-\bar{z}^2)}{2}\hat{\Theta}^{ij}_{\lambda}\Big]\mathcal{A}_{qik}^{(1)}\mathcal{A}_{\bar{q}jl}^{(1)}\epsilon^k_{\lambda}\epsilon^l_{\lambda}
	\\
	&&=2\epsilon_{\lambda}\cdot\mathcal{A}_{q}^{(0)}\epsilon_{\lambda}\cdot \mathcal{A}_{\bar{q}}^{(0)}
	+|\bm k|^2(1-\bar{z}^2)\epsilon^{i}_{\lambda}\epsilon^{j}_{\lambda}\mathcal{A}_{qil}^{(1)}\mathcal{A}_{\bar{q}j}^{(1)l},
\end{eqnarray}
\begin{eqnarray}\label{eq:rel_2}\nonumber
	2k\cdot\mathcal{A}_{q}(\bm p)k\cdot\mathcal{A}_{\bar{q}}(\bm p')
	&\rightarrow& 2|\bm k|^2\left(1-\frac{k_0^2}{\epsilon_{q\bm k}\epsilon_{\bar{q}\bm k}}\right)
	\Big[\bar{z}^2\epsilon^{i}_{\lambda}\epsilon^{j}_{\lambda}-\frac{(1-\bar{z}^2)}{2}\hat{\Theta}^{ij}_{\lambda}\Big]\mathcal{A}_{qi}^{(0)}\mathcal{A}_{\bar{q}j}^{(0)},
	\\\nonumber
	&=&-|\bm k|^2\left(1-\frac{k_0^2}{\epsilon_{q\bm k}\epsilon_{\bar{q}\bm k}}\right)\big[(1-3\bar{z}^2)\big(\epsilon_{\lambda}\cdot \mathcal{A}_{q}^{(0)}\epsilon_{\lambda}\cdot \mathcal{A}_{\bar{q}}^{(0)}\big)
	\\
	&&+(1-\bar{z}^2)\eta^{ij}\mathcal{A}_{qi}^{(0)}\mathcal{A}_{\bar{q}j}^{(0)}\big],
\end{eqnarray}
\begin{eqnarray}\label{eq:rel_3}
	{\rm Re}\big(k\cdot\epsilon(\lambda,{\bm q})\mathcal{A}_{q}(\bm p)\cdot \epsilon^*(\lambda,{\bm q})\big)k\cdot\mathcal{A}_{\bar{q}}(\bm p')
	&\rightarrow& -\bar{z}^2|\bm k|^2\epsilon_{\lambda}\cdot\mathcal{A}_{q}^{(0)}\epsilon_{\lambda}\cdot\mathcal{A}_{\bar{q}}^{(0)},
\end{eqnarray}
and
\begin{eqnarray}\label{eq:2ndrel_6}
	{\rm Re}\big(k\cdot\epsilon(\lambda,{\bm q})\mathcal{A}_{\bar{q}}(\bm p')\cdot \epsilon^*(\lambda,{\bm q})\big)k\cdot\mathcal{A}_{q}(\bm p)
	&\rightarrow& -\bar{z}^2|\bm k|^2\epsilon_{\lambda}\cdot\mathcal{A}_{q}^{(0)}\epsilon_{\lambda}\cdot\mathcal{A}_{\bar{q}}^{(0)},
\end{eqnarray}
where $\epsilon_{q/\bar{q}\bm k}\equiv\sqrt{|\bm k|^2+m^2_{q/\bar{q}}}$ is the abbreviation of $\epsilon_{q/\bar{q}}(\bm k)$.
For a complex $\epsilon^{\mu}_{\lambda}$, it is expected that one could simply replace $\big(\epsilon_{\lambda}\cdot \mathcal{A}_{q}(\bm q/2)\epsilon_{\lambda}\cdot \mathcal{A}_{\bar{q}}(\bm q/2)\big)$ by ${\rm Re}\big(\epsilon_{\lambda}\cdot \mathcal{A}_{q}(\bm q/2)\epsilon^*_{\lambda}\cdot \mathcal{A}_{\bar{q}}(\bm q/2)\big)$ and $\epsilon_{\lambda i}\mathcal{A}_{qjm}^{(1)}\epsilon^{m}_{\lambda}\mathcal{A}_{\bar{q}nl}^{(1)}$ by ${\rm Re}\big(\epsilon_{\lambda i}\mathcal{A}_{qjm}^{(1)}\epsilon^{*m}_{\lambda}\mathcal{A}_{\bar{q}nl}^{(1)}\big)$
(and similarly $\epsilon_{\lambda i}\mathcal{A}_{\bar{q}jm}^{(1)}\epsilon^{m}_{\lambda}\mathcal{A}_{qnl}^{(1)}$) in the final result.

Now, by using Eqs.~(\ref{eq:2ndrel_1})-(\ref{eq:2ndrel_6}), we find
\begin{eqnarray}\nonumber
	&&\mathcal{C}_{{\rm col},\lambda}(q)
	\\\nonumber
	&&= N_{m}\int \frac{d^3k}{(2\pi)^3}\frac{\big(M^4-(m_q^2-m_{\bar{q}}^2)^2\big)
		\delta\big(|\bm k|-\tilde{k}\big)}{8M^3\tilde{k}}
	\Bigg\{\frac{N_c^2f_{Vq}(\epsilon_{q\bm k},x_0)f_{V\bar{q}}(\epsilon_{\bar{q}\tilde{k}},x_0)}{\epsilon_{q\tilde{k}}\epsilon_{\bar{q}\tilde{k}}}\big(1-2C_m\bar{z}^2\big)
	\\\nonumber
	&&\quad-\frac{2}{m_{q}m_{\bar{q}}}\Bigg[\bigg(\big(1-2C_m\big)+\frac{\tilde{k}^2}{2\epsilon_{q\tilde{k}}\epsilon_{\bar{q}\tilde{k}}}\big(1-2C_m\bar{z}^2\big)+\frac{2k_0^2(1-\bar{z}^2)}{\epsilon_{q\tilde{k}}\epsilon_{\bar{q}\tilde{k}}}\bigg)
	\eta^{ij}\big\langle\mathcal{A}_{qi}^{a(0)}\mathcal{A}_{\bar{q}j}^{a(0)}\big\rangle
	\\\nonumber
	&&\quad+\bigg(2\big(1-C_m(1+\bar{z}^2)\big)+\frac{\tilde{k}^2}{2\epsilon_{q\tilde{k}}\epsilon_{\bar{q}\tilde{k}}}\big(1-3\bar{z}^2+2C_m\bar{z}^2(1-7\bar{z}^2)\big)+\frac{2k_0^2(1-3\bar{z}^2)}{\epsilon_{q\tilde{k}}\epsilon_{\bar{q}\tilde{k}}}\bigg) 
	\\\nonumber
	&&\quad\times{\rm Re}\Big(\big\langle\epsilon_{\lambda}\cdot\mathcal{A}_{q}^{a(0)}\epsilon^*_{\lambda}\cdot \mathcal{A}_{\bar{q}}^{a(0)}\big\rangle\Big)
	+\frac{|\bm k|^2}{2}(1-\bar{z}^2)(1-2C_m\bar{z}^2)\big\langle\mathcal{A}_{qij}^{a(1)}\mathcal{A}_{\bar{q}}^{a(1)ij}\big\rangle
	\\
	&&\quad+\frac{|\bm k|^2}{2}\big((2-5\bar{z}^2)
	-2C_m\bar{z}^2
	(1-7\bar{z}^2)
	\big)
	{\rm Re}\Big(\big\langle\epsilon^{i}_{\lambda}\epsilon^{*j}_{\lambda}\mathcal{A}_{qil}^{a(1)}\mathcal{A}_{\bar{q}j}^{a(1)l}\big\rangle\Big)
	\Bigg]
	\Bigg\},
\end{eqnarray}
where 
\begin{eqnarray}
	\epsilon_{q/\bar{q}\tilde{k}}\equiv\sqrt{\tilde{k}^2+m^2_{q/\bar{q}}}=\frac{1}{2M}\big[(m_{q/\bar{q}}^2-m_{\bar{q}/q}^2)+M^2\big]
\end{eqnarray}
and  
\begin{eqnarray}
	C_{m}\equiv\frac{\tilde{k}^2}{N_m}=\frac{1}{2}\left(1-\frac{(m_q-m_{\bar{q}})^2}{M^2}\right).
\end{eqnarray}
Here we have presumably evaluated the correlation of Wigner functions for the quark and antiquark before conducting the $k$ integral and written down a generic expression for both real and complex $\epsilon_{\lambda}$. 
After integrating over $\bar{z}$, we arrive at Eq.~(\ref{eq:f_lambda_general})
\bibliography{spin_alignment_color_main.bbl}
\end{document}